\documentclass[final,3p,times,twocolumn,citesort]{elsarticle}
\usepackage{epsfig}
\usepackage{amssymb} 

\journal{Computational Materials Science}
\def\ffrac#1#2{ \hspace{3pt}\!^{#1}\!\!\hspace{1pt}/ \hspace{2pt}\!\!_{#2}\!\hspace{3pt} }
\def\half{{\ffrac{1}{2}}}
\def\atrd{{\ffrac{1}{3}}}
\def\afth{{\ffrac{1}{4}}}
\def\tfth{{\ffrac{3}{4}}}

\def\teth{{\ffrac{3}{8}}}
\def\feth{{\ffrac{5}{8}}}
 \def\kk{{\bf k}}
\def\aa{{{\bf a}}} \def\bb{{{\bf b}}}

\begin{document}

\begin{frontmatter}
  \title{High-throughput electronic band structure calculations: challenges and tools}
  \author{Wahyu Setyawan$^1$ and Stefano Curtarolo$^{1,2}$}
  \address{
    $^1$Dept. of Mechanical Engineering and Materials Science and Dept. of Physics, Duke University, Durham, NC 27708.\\
    $^2$corresponding author: stefano@duke.edu
  }
  
  \begin{abstract}
The article is devoted to the discussion of the high-throughput
approach to band structures calculations. We present scientific and
computational challenges as well as solutions relying on the developed
framework (Automatic Flow, {\small AFLOW/ACONVASP}). The key factors of the
method are the standardization and the robustness of the procedures.
Two scenarios are relevant: 1) independent users generating databases
in their own computational systems (off-line approach) and 2) teamed
users sharing computational information based on a common ground
(on-line approach). Both cases are integrated in the framework: for
off-line approaches, the standardization is automatic and fully
integrated for the 14 Bravais lattices, the primitive and conventional
unit cells, and the coordinates of the high symmetry $\kk$-path in the
Brillouin zones. For on-line tasks, the framework offers an expandable
web interface where the user can prepare and set up calculations
following the proposed standard. Few examples of band structures are
included. LSDA+U parameters ($U, J$) are also presented for Nd, Sm, and
Eu.  
  \end{abstract}
  
\end{frontmatter}

Over the past decade computational materials science has undergone a tremendous growth 
thanks to the availability, power and relatively limited cost of high-performance computational equipment.
The high-throughput (HT) method,
started from the seminal paper by Xiang {\it et al.} for combinatorial discovery of superconductors \cite{Xiang06231995},
has become an effective and efficient tool for materials development
\cite{koinuma_nmat_review2004,Spivack20035,Boussie2003,Potyrailo2003,Potyrailo2005}
and prediction
\cite{ceder:nature_1998,Johann02,Stucke03,curtarolo:prl_2003_datamining,morgan:meas_2005_ht,monster,Fischer06}.
Recent examples of computational HT are 
the ‘‘Pareto-optimal” search for alloys and catalysts \cite{johannesson:ref2,johannesson:ref3},
the ‘‘data-mining of quantum calculations” method leading to the principle-component analysis of the formation energies 
of many alloys in several configurations \cite{curtarolo:prl_2003_datamining,morgan:meas_2005_ht,monster,curtarolo:art53,curtarolo:art49},
the ‘‘high-throughput Kohn-anomalies” search in ternary 
lithium-borides \cite{curtarolo:art21,calandra:LiB_superconductivity_2007,kolmogorov:ternary_borides_LiB_2008},
and the ‘‘multi-optimization” techniques used for the study of 
high-temperature reactions in multicomponent hydrides \cite{Wolverton2008,Siegel_PhysRevB.76.134102,Akbarzadeh2007}.

In its practical implementation, HT uses some sort of automatic optimization technique to screen 
through a library of candidate compounds and to direct further refinements.
The library can be a set of alloy prototypes \cite{Massalski,monster}
or a database of compounds such as the Pauling File \cite{Pauling} or the ICSD Database \cite{ICSD,ICSD3}. 
An important difference between the ‘‘several-calculations” and the ‘‘HT” 
philosophies is that the former concentrates on the calculation of a particular property, 
while the latter focuses on the extraction of property correlations which are used to guide the search for systems with {\it ad-hoc} characteristics.
The power of HT comes with a cost. Due to the enormous amount of information produced, standardization and robustness of the procedures are necessary.
This is especially true if one were concomitantly optimizing thermodynamics and electronic structure, which is required, for instance, 
in catalyst design \cite{Norskov2009natchem}, 
in accelerated ``battery materials'' discovery \cite{Ceder_Chem_Materials},
and superconducting materials development \cite{calandra:LiB_superconductivity_2007,kolmogorov:ternary_borides_LiB_2008}.
Therefore a rational HT computational framework must contain a general, reliable, and standardized electronic structure 
analysis feature. It must determine the symmetry automatically, the Brillouin zone (BZ) integration path for all the possible 
14 Bravais Lattices with all their various sub-cases, and put the direct and reciprocal lattice vectors in the appropriate
standardized form, so that data can be exchanged and recycled between different projects. Although Brillouin zones integration paths 
have been included in books and literature for the last few decades \cite{Bradley72,Burns90,Miller67,Kovalev65,Casher69},
a standardized definition of the paths for all the different cases is, 
to the best of our knowledge, missing. 

In this article, we describe the BZ paths features of {\small AFLOW} \cite{AFLOW}, 
which is our free framework for performing high-throughput thermodynamics and 
electronic structure calculations on top of DFT ab initio codes 
(currently the Vienna Ab-initio Simulation Package ({\small VASP}) but the porting to other DFT packages, such  as Quantum Espresso \cite{quantum_espresso_2009} is underway).
A typical task involves structural optimizations to a ground state (``relaxation run''), 
determination of charge density and its projection onto electronic orbitals (``static run''),
and calculation of energy levels along a path of ``important'' wave vectors (``bands run'').
We refer this set of $\kk$-points as $\kk$-path.

\begin{figure}[h!]
\centerline{\epsfig{file=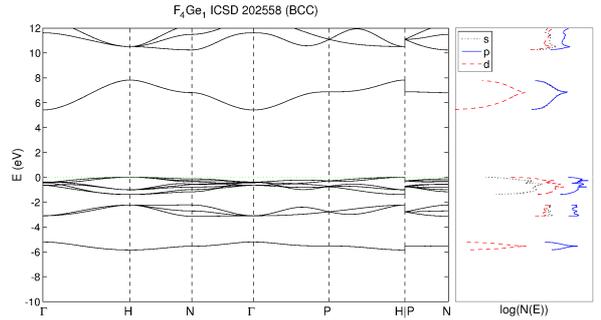
,height=42mm}}
\vspace{-3mm}
\caption{\small
Band structure of GeF$_4$ in BCC lattice.}
\label{figbandBCC}
\end{figure}

Some definitions are pertinent for automatic construction of the $\kk$-path.
A $\kk$-point is a {\it symmetry point} if its site symmetry contains at least one point symmetry operation 
that does not belong to the site symmetry of the neighboring points in sufficiently small vicinity.
Similarly, a line or a plane forms a {\it symmetry line} or a {\it symmetry plane} if it contains at least one symmetry point 
and all of the $\kk$-points on the line or plane have site symmetry with at least one point symmetry operation not possessed by the neighboring points.
The $\kk$-path must be carefully chosen so that the electronic properties of a materials imposed by its underlying crystal symmetry
are correctly described.
For example, GeF$_4$ (cI10, ICSD \#202558, space group \#217, I$\bar{4}$3m) has an indirect gap. 
The conduction band minimum (CBM) occurs at point $\Gamma$, while the valence band maximum (VBM) occurs at the point H of the BZ 
for the body-centered cubic (BCC) lattice (Figure \ref{figbandBCC}). 
If a cubic unit cell were used instead the BCC, one would incorrectly find a direct gap occurring at point $\Gamma$.

The coordinates of symmetry $\kk$-points are more conveniently expressed as fractions of the reciprocal lattice vectors.
Therefore, the primitive lattice vectors needs to be properly defined in a standardized fashion.
In order to build a primitive cell, the framework {\small AFLOW} performs the following procedure. \\
{\bf i.}
Given any input structure (unit cell and the coordinates of the basis atoms), {\small AFLOW} reduces it into a minimal set of basis atoms in a primitive cell. \\
{\bf ii.}
A set of symmetry properties are calculated: 
lattice point group, crystal point group, crystal family, factor group, space group operations, Pearson symbol, and the Bravais lattice type. 
If the input structure contains information about the space group number, it will be used to double check the Bravais lattice type. \\
{\bf iii.}
A {\it standard conventional} cell is then identified and constructed, and whenever possible, the lattice vectors are ordered according to the axial lengths 
and the interaxial angles. This ordering eliminates some choices in the possible shapes of BZ in certain Bravais lattices. 
For example, let us consider the body-centered orthorhombic lattice.
Depending on the ratios of the axial lengths the body-centered orthorhombic cell has three possible shapes of the BZ, 
and consequently three different coordinates of the symmetry points. 
By ordering the conventional lattice vectors so that $|\aa_1|<|\aa_2|<|\aa_3|$ 
(i.e. $a<b<c$) a unique choice remains. Furthermore, the ordering improves the similarity between band 
structures of different compounds with the same lattice: the proportion between the length of each path
segment in the band structure will be comparable. \\
{\bf iv.}
From the ordered conventional unit cell, a {\it standard primitive} cell is created. 
Amongst all the possible primitive cells, {\small AFLOW}, chooses the one with the reciprocal lattice vectors
passing through the center of the Bragg planes belonging to the first BZ.
The choice has considerable practical advantages. It enforces the reciprocal lattice vectors to be as 
perpendicular as possible within each other (Minkowski lattice reduction \cite{Nguyen2009_Minkowsky}) 
and minimizes the number of the plane waves basis set used in the quantum mechanical code, allowing faster
convergence and smaller memory requirements.
\footnote{The common visualization softwares used for generating BZs simply take the input lattice vectors,
  generate reciprocal vectors and produce BZs which are not necessarily the Wigner-Seitz cells of the reciprocal lattice.}
After all these steps, the standard primitive unit cells of the 14 Bravais lattices as calculated by the {\small AFLOW} package 
are safe to be used in the ``relaxation'' and ``static'' calculations. As usual, the latter calculation will be used for the electronic density of states.

\begin{figure}[h!]
\centerline{\epsfig{file=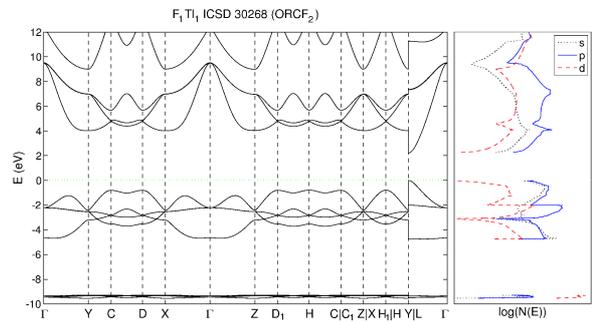
,height=42mm}}
\vspace{-3mm}
\caption{\small
Band structure of TlF in ORCF$_2$ lattice.}
\label{figbandORCF2}
\end{figure}

The $\kk$-paths used in the band structure analysis are constructed from the irreducible part of the first Brillouin zone (IRBZ). 
A symmetry line is included in the path if it belongs to the edges of the IRBZ, otherwise it is included only if it carries one or 
more new point symmetry operations with respect to those of its extremes.
Duplicate lines, due to the reciprocal lattice point group symmetry and translations are also omitted.
A point possessing only identity ($E$) and inversion ($I$) operations can not form a symmetry line, however, it may still be a zero-slope point \cite{Cracknell73}. Examples are point L in face-centered orthorhombic, points N, N$_1$, and M in C-centered monoclinic, and all symmetry points in triclinic. To illustrate, TlF (oF8, ICSD \#30268, space group \#69, Fmmm) has CBM and VBM occuring at point L (Figure \ref{figbandORCF2}). This result would have not been obtained if point L were excluded. Therefore, for completeness, a line from point $\Gamma$ to such point is included in the $\kk$-path.

\begin{figure}[hbp!]
\centerline{\epsfig{file=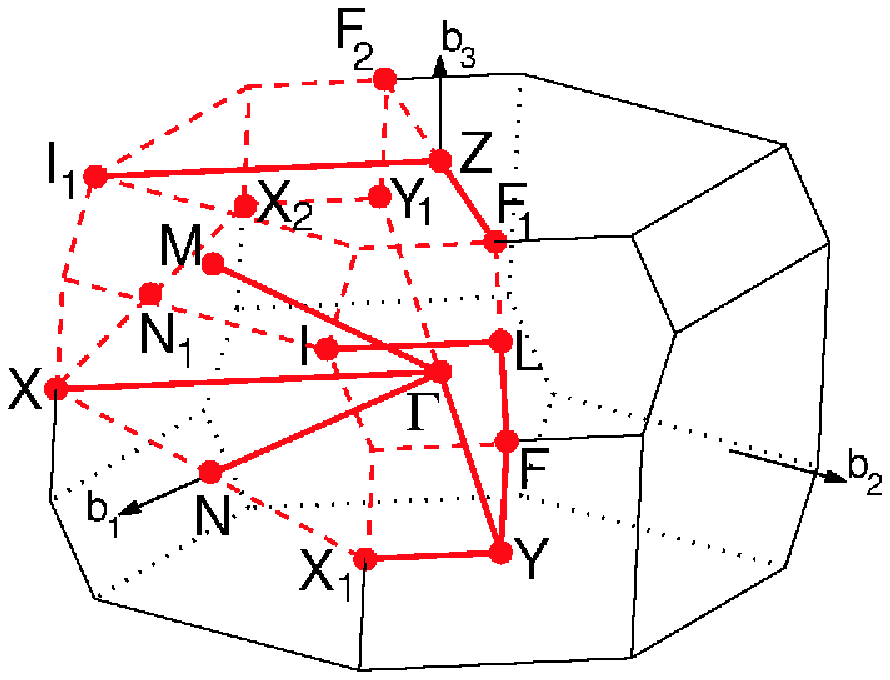
,width=55mm}}
\vspace{-1mm}
\caption{\small 
Brillouin zone of MCLC$_1$ lattice. Path: $\Gamma$-Y-F-L-I$|$I$_1$-Z-F$_1|$Y-X$_1|$X-$\Gamma$-N$|$M-$\Gamma$. An example of band structure using this path is given in Figure \ref{figbandMCLC1}.}
\label{figMCLC1}
\end{figure}

Notably for triclinic, monoclinic, and rhombohedral systems, the shape of Wigner-Seitz cell of the reciprocal lattice
depends nontrivially on the lattice vectors. For this reason, some researchers use the parallelepiped of primitive
reciprocal lattice vectors, centered at $\kk=$ {\bf 0}, to define a plausible BZ.
Even though the energy is continuous throughout such BZ
\cite{Bradley72}, 
its faces, in general, are not parallel to the symmetry planes of the lattice. 
Therefore, there is no guarantee that any line connecting two $\kk$-points
on a BZ face will be a symmetry line. In addition, since parallelepiped unit cells have only 8 points, one at each corner of the IRBZ, 
one would miss some symmetry points on the Wigner-Seitz cell of reciprocal lattice.
Consequently, a complete irreducible set of symmetry lines would not be obtained.
For example if we were using a parallelepiped as BZ in the C-centered monoclinic variation MCLC$_1$ (Figure \ref{figMCLC1}),
the site symmetries of points $\ffrac{1}{2}\bb_1$, $\ffrac{1}{2}\bb_2$, $\ffrac{1}{2}(\bb_1+\bb_3)$, and $\ffrac{1}{2}(\bb_2+\bb_3)$ would be only $E$ and $I$. 
Therefore, the line $\ffrac{1}{2}\bb_1$-Y would not be, for example, a symmetry line. 
Furthermore, points like X, X$_1$, I, and I$_1$ and their related symmetry lines ($C2$ about $x$-axis) would not be included in such simplification.
To conclude, although we believe that the simple parallelepiped BZ can be useful in some particular difficult cases,
we think that the solution is not appropriate for a full automatic and high-throughput implementation of the band structure analysis.
For this reason, inside the framework {\small AFLOW}, all the BZ and their $\kk$-paths are derived from the Wigner-Seitz cell of the reciprocal lattice and the available symmetries.
In Appendix, for all the Bravais lattices and variations,
we present the conventional and primitive lattice vectors implemented in {\small AFLOW},
the coordinates of high-symmetry $\kk$-points for the path, the shape of the BZ, and an example of band structure calculation for a compound extracted from the ICSD database. \\
{\bf Off-line implementation.}

The effort in developing the standardized tool {\small AFLOW} comes from the ongoing generation of an extensive
database of electronic band structure for inorganic crystals for scintillator materials design \cite{curtarolo:art46}.
We have extracted approximately 195,000 structures from the
Inorganic Crystal Structure Database (ICSD) \cite{ICSD,ICSD1,ICSD2,ICSD3}. {\small AFLOW} is equipped with utilities to select structures of interest. 
Selection criteria can be based on atomic number or element's name, mass density, number of atoms per primitive unit cell, chemical formula, structure prototype, 
space group number, lattice type, ICSD entry number, etc. Features to remove/include structures containing certain elements, partial occupancies, and redundancies
(structures with the same chemical formula and space group number) are also implemented. 
Each structure is given a label which is composed of the structure`s chemical formula (in alphabetic order) and the ICSD entry number.
This label is the only information that {\small AFLOW} requires to produce the band structure. 
After the structure selection is performed, a list of labels is produced. Based on such labels, {\small AFLOW} creates a subdirectories 
for each structure and the necessary input file for the band structure calculation with {\small VASP} (porting to other DFT packages, such as Quantum Espresso \cite{quantum_espresso_2009} is underway).
For running the DFT package, {\small AFLOW} has an option to run only one structure and exit, or to search through subfolders and run those that have not been calculated yet, 
or to wait for new structures to run. If started as a common Unix daemon through the queue of a computer cluster, {\small AFLOW} will generate, run, correct and converge many
calculations per day, with minimum human input.\\
{\bf Electronic structure database implementation.}

The database of band structures under construction is calculated using {\small VASP} within the General Gradient Approximation of the density functional theory \cite{DFT}. 
We use projector augmented waves pseudopotentials with exchange correlation functionals as parameterized by Perdew-Burke-Ernzerhof \cite{PAW,PBE}. 
All structures are fully relaxed with a convergence tolerance of 1 meV/atom using dense grids of 3,000-4,000 $\kk$-points per reciprocal atom for the integrations over BZ. 
A much denser grid of 10,000 is implemented for the static run to get accurate charge densities and density of states.
Monkhorst-Pack scheme \cite{MonkhorstPack} is employed in the grid generation except for hexagonal and rhombohedral systems in which $\Gamma$-centered grid 
is selected for faster convergence. At the beginning of relaxation, a spin-polarized calculation is performed for all structures.
Then, if the magnetization is smaller than 0.025 $\mu_B$/atom, the spin is turned off for the next relaxations and subsequent calculations to enhance the calculation speed.
At the completion of each calculation (``relax''$\rightarrow$``relax''$\rightarrow$``static''$\rightarrow$``bands'''), appropriate {\small MATLAB}
scripts are invoked for data analysis and visualization. All these steps are done automatically.
One of the most difficult challenges in the high-throughput combinatorial search is about the response to erroneous interruption 
of the one or more of the flows working on a big set of problems, concurrently.
The most common cause is insufficient hardware resources. 
Precaution must be taken, for example by estimating the memory requirement of the tasks, 
by grouping jobs based on memory, and by adapting the number of concurrent allocated CPUs
with respect to the expected simulation speed.
In addition, in many shared high-performance computer facilities, walltime is limited. 
This imposes a difficult problem because estimating computer time {\it a priori} is highly nontrivial, 
especially since 
the number of the required electronic and ionic relaxations 
depends on how distant the initial configuration is from the unknown final equilibrium.
The second most common cause of interruption is due to runtime errors of {\small VASP}. 
{\small AFLOW} is capable of detecting most of the problems and it contains many self healing features.
This is achieved by diagnosing the error message, self-correcting the appropriate parameters, and restarting {\small VASP}.
With {\small AFLOW}, a job can be easily restarted from ``relaxations'', ``static'', or ``bands'' steps.
{\small AFLOW}'s capability to continuously search and manage sub folders are not limited to DFT calculation. 
An {\it alien} mode is implemented, which allows {\small AFLOW} to execute other tasks in a high-throughput fashion. 
For instance, the many thousands Grand Canonical Monte Carlo calculations used 
in a recent surface science absorption project \cite{curtarolo:art44,curtarolo:art43},
were directed and performed by {\small AFLOW}.
In addition, {\small AFLOW} is equipped with options to run a "pre" and "post" command/scripts 
that will be executed before and after the main program is performed in each folder, respectively.
This allows {\small AFLOW} to generate input files on the fly depending on the results of different calculations, so that ad-hoc optimization can be implemented by the users.
In conclusion, the ``alien'' mode and the ``pre/post'' command options improves the flexibility on the recovery from a
crash or an unconverged run in a high-throughput manner as well as increases the overall versatility and throughput of {\small AFLOW}. \\
{\bf Implemented electronic properties.}

A typical information that one can extract from the band structure calculations includes the Fermi energy, band gap, type of the band gap, 
width of valence and conduction bands, effective mass of electron and hole, charge densities, band structures, total and partial density of states, etc. 
A user can easily create utilities in any language at choice for data analysis,
and use {\small AFLOW} in ``alien'' mode to execute the utilities automatically in each subfolders. 
For our purpose we have chosen {\small MATLAB}, which has produced all the band structure and orbital-projected total density of states for every shape the BZ as presented in the Appendix. \\
{\bf LSDA+U corrections.}

It is generally known that due to a rather weak orbital-dependence of the DFT's exchange correlation energy, 
the strong on-site Coulomb repulsion in systems with narrow d- and f-bands is underaccounted.
As a result, DFT produces bandgaps that are smaller than experimental values and some times it fails
to get the correct ground state in such systems. 
Based on our experience in calculating the electronic structure of many lanthanum halides, DFT incorrectly gives
conduction band minima with 4f states instead of 5d orbitals. The insufficient description of strongly correlated systems given by DFT,
can be remedied, at least partially, by GW \cite{GW} or LSDA+U corrections.
Due to their large computational cost, GW corrections are not currently applicable for high-throughput searches.
When needed, LSDA+U corrections are automatically implemented by {\small AFLOW}, based on the formulas developed by 
Duradev \cite{DuradevDFTU} and Liechtenstein \cite{LiechDFTU}.
To the best of our knowledge, there are no systematic studies for the Hubbard $U$ and the screened Stoner exchange parameters $J$ 
across all the elements with all the possible oxidation states. 
Detailed analysis and determination of the $U$ and $J$ values for different compounds
would be one of the tasks of high-throughput future research \cite{Cococcioni_JPCM}.
In the mean time, we have applied the $+U$ corrections to the 4f-wave functions of lanthanide compounds
to get the correct orbitals at the conduction band minimum.
For the systems not available in literature but related to our current research, Nd, Sm, and Eu,
we have fit $U$ and $J$ so that the 4f levels reproduce the experimental density of states from the x-ray photoelectron spectroscopy and 
Bremsstrahlung isochromat spectroscopy (XPS-BIS) measurements \cite{Lang81}. 
Although the data is for metals, we are confident that the values of $U$ and $J$ for other compounds will not be very different 
from the fit. The values are listed in table \ref{table_U}.

\begin{table}
\caption{\small 
  Default value of $U$ and $J$ parameters given in eV applied to f-orbitals
  within the GGA+U approximation included in {\small AFLOW}. 
  Note that these parameters are subject to update.}
\begin{tabular}{lccl|lccl}
\hline \hline
atom & $U$ & $J$ & ref. & atom & $U$ & $J$ & ref.\\
\hline
La & 8.1 & 0.6 & \cite{LaUJ} 	& Eu & 6.4 & 1.0 & \\
Ce & 7.0 & 0.7 & \cite{CeUJ} 	& Gd & 6.7 & 0.7 & \cite{GdUJ} \\
Pr & 6.5 & 1.0 & \cite{PrUJ}	& Tm & 7.0 & 1.0 & \cite{TmUJ} \\
Nd & 7.2 & 1.0 & 		& Yb & 7.0 & 0.67 & \cite{YbUJ} \\
Sm & 7.4 & 1.0 & 	      	& Lu & 4.8 & 0.95 & \cite{LaUJ} \\
\hline
\end{tabular}
\label{table_U}
\end{table}
\noindent {\bf On-line implementation: aconvasp-online.}

Users who do not need to perform high-throughput calculations or to create databases, 
can prepare standard unit cells input files and extract the appropriate $\kk$-points path 
by using the command version of {\small AFLOW} called {\small ACONVASP} or 
the on-line tool {\small ACONVASP}-online available in our website (http://materials.duke.edu).

The following protocol should be followed.
Unit cells must first be reduced to standard primitives, then they should be appropriately relaxed (if needed).
Before the static run, the cells should be reduced again to standard primitive (symmetry and orientation 
might have changed during the relaxation). 
The user should then perform the static run and then project the eigenvalues along the directions which are specified in the ``kpath'' option.
If the user is running {\small AFLOW} and {\small VASP}, 
the web interface can also prepare a template input file ``aflow.in'' which performs all the mentioned tasks.

\begin{center}
---------------
\end{center}

\section{Appendix A}
\label{appendixA}
\noindent
The choice of lattice vectors implemented in {\small AFLOW} is given here. When the primitive lattice is the same as the conventional one, it is simply called ``lattice''. Variables $a,b,c,\alpha,\beta,\gamma$ denote the axial lengths and interaxial angles of the conventional lattice vectors, while $k_a,k_b,k_c,k_\alpha,k_\beta,k_\gamma$ are those of the primitive reciprocal lattice vectors $\bb_1,\bb_2,\bb_3$. The coordinates of symmetry $\kk$-points are given in fractions of $\bb_1,\bb_2,\bb_3$.\\

\subsection{Cubic (CUB, cP)}
\noindent
Lattice:\\
$\aa_1 = (a, 0, 0)$\\
$\aa_2 = (0, a, 0)$\\
$\aa_3 = (0, 0, a)$\\
\vspace{-10mm}
\begin{center}
\begin{table}[hbp!]
\caption{Symmetry $\kk$-points of CUB lattice.}
\begin{tabular}{cccl|cccl}
\hline \hline
$\times \bb_1$ & $\times \bb_2$ & $\times \bb_3$ & & $\times \bb_1$ & $\times \bb_2$ & $\times \bb_3$ &\\
\hline
   0 &  0 & 0 & $\Gamma$&   $\half$ &	$\half$ & $\half$ & R\\
   $\half$ &	$\half$ & 0 & M&   0 &	$\half$ & 0 & X\\
\hline
\end{tabular}
\end{table}
\end{center}
\vspace{-10mm}
\begin{figure}[hbp!]
\centerline{\epsfig{file=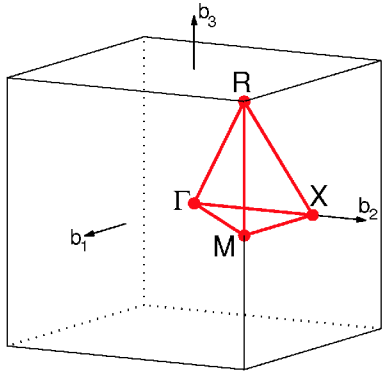,width=50mm}}
\vspace{-1mm}
\caption{\small 
Brillouin zone of CUB lattice. Path: $\Gamma$-X-M-$\Gamma$-R-X$|$M-R. An example of band structure using this path is given in Figure \ref{figbandCUB}.}
\label{figCUB}
\end{figure}

\subsection{Face-centered Cubic (FCC, cF)}
\noindent
\begin{tabular}{l|l}
Conventional lattice: & Primitive lattice:\\
$\aa_1 = (a, 0, 0)$ & $\aa_1 = (0  , a/2, a/2)$\\
$\aa_2 = (0, a, 0)$ & $\aa_2 = (a/2, 0  , a/2)$\\
$\aa_3 = (0, 0, a)$ & $\aa_3 = (a/2, a/2, 0  )$\\
\end{tabular}

\vspace{-5mm}
\begin{center}
\begin{table}[htp!]
\caption{Symmetry $\kk$-points of FCC lattice.}
\begin{tabular}{cccl|cccl}
\hline \hline
$\times \bb_1$ & $\times \bb_2$ & $\times \bb_3$ & & $\times \bb_1$ & $\times \bb_2$ & $\times \bb_3$ &\\
\hline
   0 &  0 &  0 &  $\Gamma$&   $\feth$ &  $\afth$ &  $\feth$ &  U \\
   $\teth$ &  $\teth$ &  $\tfth$ &  K&   $\half$ &  $\afth$ &  $\tfth$ &  W\\
   $\half$ &  $\half$ &  $\half$ &  L&   $\half$ &  0 &  $\half$ &  X\\
\hline
\end{tabular}
\end{table}
\end{center}

\vspace{-8mm}
\begin{figure}[htp!]
\centerline{\epsfig{file=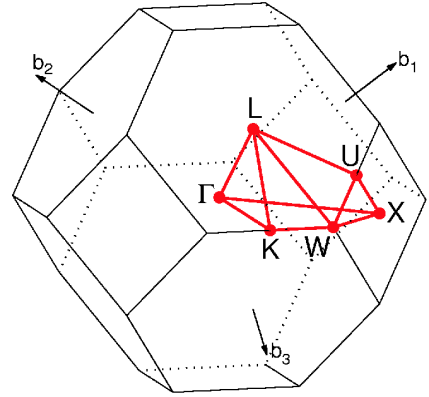,width=55mm}}
\vspace{-1mm}
\caption{\small 
Brillouin zone of FCC lattice. Path: $\Gamma$-X-W-K-$\Gamma$-L-U-W-L-K$|$U-X. An example of band structure using this path is given in Figure \ref{figbandFCC}.}
\label{figFCC}
\end{figure}

\subsection{Body-centered Cubic (BCC, cI)}
\noindent
\begin{tabular}{l|l}
Conventional lattice: & Primitive lattice:\\
$\aa_1 = (a, 0, 0)$ & $\aa_1 = (-a/2,  a/2,  a/2)$\\
$\aa_2 = (0, a, 0)$ & $\aa_2 = (a/2,  -a/2,  a/2)$\\
$\aa_3 = (0, 0, a)$ & $\aa_3 = (a/2,  a/2,  -a/2)$\\
\end{tabular}

\vspace{-8mm}
\begin{center}
\begin{table}[hbp!]
\caption{Symmetry $\kk$-points of BCC lattice.}
\begin{tabular}{cccl|cccl}
\hline \hline
$\times \bb_1$ & $\times \bb_2$ & $\times \bb_3$ & & $\times \bb_1$ & $\times \bb_2$ & $\times \bb_3$ &\\
\hline
   0 &  0 &  0 & $\Gamma$&   $\afth$ &  $\afth$ &  $\afth$ & P \\
   $\half$ & -$\half$ &  $\half$ & H&    0 &  0 &  $\half$ & N \\
\hline
\end{tabular}
\end{table}
\end{center}

\vspace{-8mm}
\begin{figure}[hbp!]
\centerline{\epsfig{file=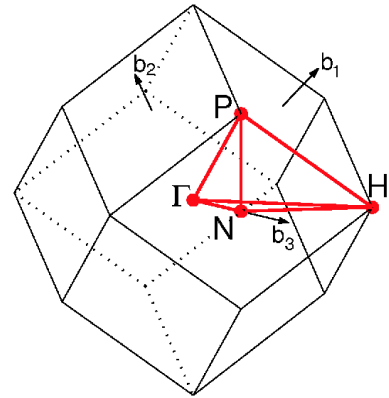,width=50mm}}
\vspace{-1mm}
\caption{\small 
Brillouin zone of BCC lattice. Path: $\Gamma$-H-N-$\Gamma$-P-H$|$P-N. An example of band structure using this path is given in Figure \ref{figbandBCC}.}
\label{figBCC}
\end{figure}

\pagebreak
\subsection{Tetragonal (TET, tP)}
\noindent
Lattice:\\
$\aa_1 = (a, 0, 0)$\\
$\aa_2 = (0, a, 0)$\\
$\aa_3 = (0, 0, c)$\\

\vspace{-5mm}
\begin{center}
\begin{table}[hbp!]
\caption{Symmetry $\kk$-points of TET lattice.}
\begin{tabular}{cccl|cccl}
\hline \hline
$\times \bb_1$ & $\times \bb_2$ & $\times \bb_3$ & & $\times \bb_1$ & $\times \bb_2$ & $\times \bb_3$ &\\
\hline
   0 &  0 &  0 &  $\Gamma$&   0 &  $\half$ &  $\half$ &  R\\
   $\half$ &  $\half$ &  $\half$ &  A&   0 &  $\half$ &  0 &  X\\
   $\half$ &  $\half$ &  0 &  M&   0 &  0 &  $\half$ &  Z\\
\hline
\end{tabular}
\end{table}
\end{center}

\vspace{-5mm}
\begin{figure}[hbp!]
\centerline{\epsfig{file=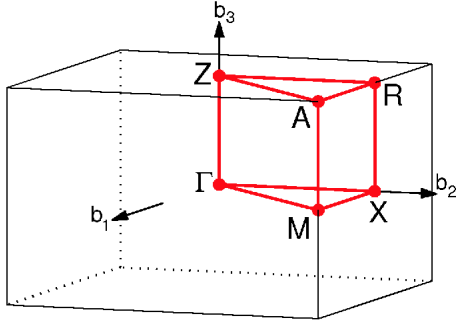,width=60mm}}
\vspace{-1mm}
\caption{\small 
Brillouin zone of TET lattice. Path: $\Gamma$-X-M-$\Gamma$-Z-R-A-Z$|$X-R$|$M-A. An example of band structure using this path is given in Figure \ref{figbandTET}.}
\label{figTET}
\end{figure}

\subsection{Body-centered Tetragonal (BCT, tI)}
\noindent
\begin{tabular}{l|l}
Conventional lattice: & Primitive lattice:\\
$\aa_1 = (a, 0, 0)$ & $\aa_1 = (-a/2,  a/2,  c/2)$\\
$\aa_2 = (0, a, 0)$ & $\aa_2 = (a/2,  -a/2,  c/2)$\\
$\aa_3 = (0, 0, c)$ & $\aa_3 = (a/2,  a/2,  -c/2)$\\
\end{tabular}

\noindent
Variations:\\
BCT$_1$: $c<a$\\
BCT$_2$: $c>a$\\

\vspace{-5mm}
\begin{center}
\begin{table}[hbp!]
\caption{Symmetry $\kk$-points of BCT$_1$ lattice.}
\begin{tabular}{cccl|cccl}
\hline \hline
$\times \bb_1$ & $\times \bb_2$ & $\times \bb_3$ & & $\times \bb_1$ & $\times \bb_2$ & $\times \bb_3$ &\\
\hline
   0 &  0 &  0 &  $\Gamma$&   0 &  0 &  $\half$ &  X\\
  -$\half$ &  $\half$ &  $\half$ &  M&   $\eta$   &  $\eta$   & -$\eta$   &  Z\\
   0 &  $\half$ &  0 &  N&  -$\eta$   &  1-$\eta$ &  $\eta$   &  Z$_1$\\
   $\afth$ &  $\afth$ &  $\afth$ &  P\\
\vspace{-3mm}\\
\multicolumn{8}{l}{$\eta=(1+c^2/a^2)/4$}\\
\hline
\end{tabular}
\end{table}
\end{center}

\begin{figure}[htp!]
\centerline{\epsfig{file=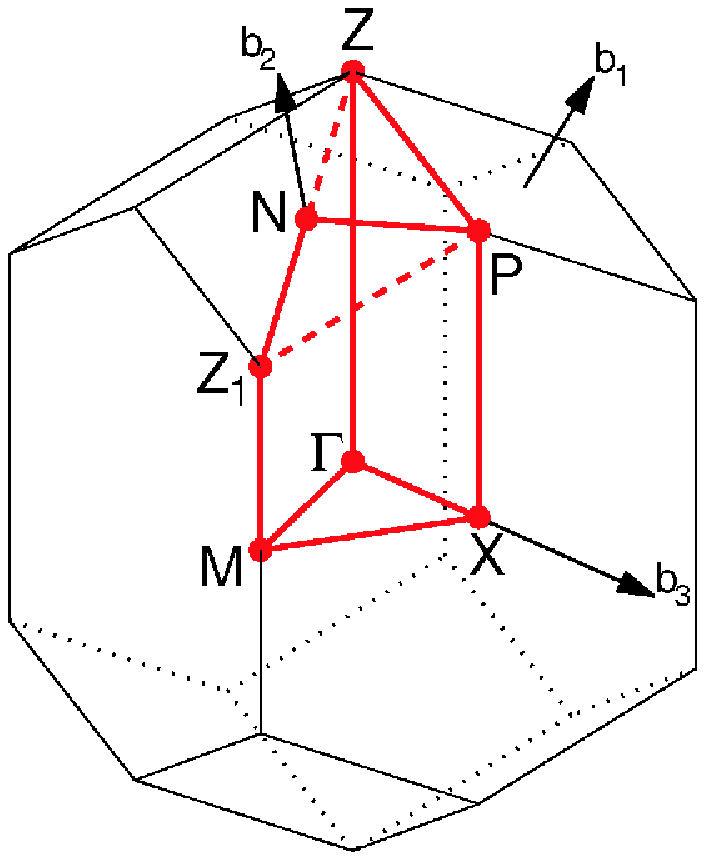,width=45mm}}
\vspace{-1mm}
\caption{\small 
Brillouin zone of BCT$_1$ lattice. Path: $\Gamma$-X-M-$\Gamma$-Z-P-N-Z$_1$-M$|$X-P. An example of band structure using this path is given in Figure \ref{figbandBCT1}.}
\label{figBCT1}
\end{figure}

\vspace{-5mm}
\begin{center}
\begin{table}[hbp!]
\caption{Symmetry $\kk$-points of BCT$_2$ lattice.}
\begin{tabular}{cccl|cccl}
\hline \hline
$\times \bb_1$ & $\times \bb_2$ & $\times \bb_3$ & & $\times \bb_1$ & $\times \bb_2$ & $\times \bb_3$ &\\
\hline
   0 &  0 &  0 &  $\Gamma$                  &  0 &  0 &  $\half$ &  X\\
   0 &  $\half$ &  0 &  N      &   -$\zeta$  &  $\zeta$  &  $\half$ &  Y\\
   $\afth$ &  $\afth$ &  $\afth$ &  P                   &  $\half$ &  $\half$ & -$\zeta$  &  Y$_1$\\
   -$\eta$   &  $\eta$   &  $\eta$   &  $\Sigma$       &   $\half$ &  $\half$ & -$\half$ &  Z\\
   $\eta$   &  1-$\eta$ & -$\eta$   &  $\Sigma_1$\\
\vspace{-3mm}\\
\multicolumn{4}{l}{$\eta=(1+a^2/c^2)/4$}, & \multicolumn{4}{l}{$\zeta=a^2/(2c^2)$}\\
\hline
\end{tabular}
\end{table}
\end{center}

\vspace{-5mm}
\begin{figure}[hbp!]
\centerline{\epsfig{file=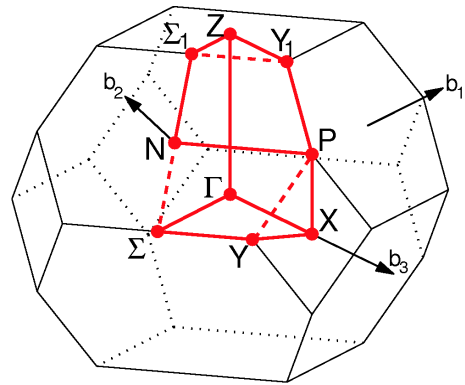,width=60mm}}
\vspace{-1mm}
\caption{\small 
Brillouin zone of BCT$_2$ lattice. Path: $\Gamma$-X-Y-$\Sigma$-$\Gamma$-Z-$\Sigma_1$-N-P-Y$_1$-Z$|$X-P. An example of band structure using this path is given in Figure \ref{figbandBCT2}.}
\label{figBCT2}
\end{figure}

\pagebreak
\subsection{Orthorhombic (ORC, oP)}
\noindent
Ordering of the conventional lattice: $a<b<c$. Lattice:\\
$\aa_1 = (a, 0, 0)$\\
$\aa_2 = (0, b, 0)$\\
$\aa_3 = (0, 0, c)$\\

\vspace{-10mm}
\begin{center}
\begin{table}[hbp!]
\caption{\small Symmetry $\kk$-points of ORC.}
\begin{tabular}{cccl|cccl}
\hline \hline 
$\times \bb_1$ & $\times \bb_2$ & $\times \bb_3$ & & $\times \bb_1$ & $\times \bb_2$ & $\times \bb_3$ &\\
\hline 
   0 &  0 & 0 & $\Gamma$&   $\half$ &  0 & $\half$ & U\\
   $\half$ &  $\half$ & $\half$ & R&   $\half$ &  0 & 0 & X\\
   $\half$ &  $\half$ & 0 & S&   0 &  $\half$ & 0 & Y\\
   0 &  $\half$ & $\half$ & T&   0 &  0 & $\half$ & Z\\
\hline
\end{tabular}
\end{table}
\end{center}

\vspace{-10mm}
\begin{figure}[hbp!]
\centerline{\epsfig{file=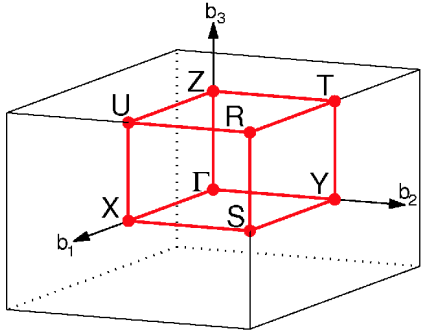,width=55mm}}
\vspace{-1mm}
\caption{\small 
Brillouin zone of ORC lattice. Path: $\Gamma$-X-S-Y-$\Gamma$-Z-U-R-T-Z$|$Y-T$|$U-X$|$S-R. An example of band structure using this path is given in Figure \ref{figbandORC}.}
\label{figORC}
\end{figure}

\subsection{Face-centered Orthorhombic (ORCF, oF)}
\noindent
Ordering of the conventional lattice: $a<b<c$.\\
\begin{tabular}{l|l}
Conventional lattice: & Primitive lattice:\\
$\aa_1 = (a, 0, 0)$ & $\aa_1 = (0  , b/2, c/2)$\\
$\aa_2 = (0, b, 0)$ & $\aa_2 = (a/2, 0  , c/2)$\\
$\aa_3 = (0, 0, c)$ & $\aa_3 = (a/2, b/2, 0  )$\\
\end{tabular}

\noindent
Variations:\\
ORCF$_1$:  $1/a^2 > 1/b^2 + 1/c^2$\\
ORCF$_2$:  $1/a^2 < 1/b^2 + 1/c^2$\\
ORCF$_3$:  $1/a^2 = 1/b^2 + 1/c^2$\\

\vspace{-10mm}
\begin{center}
\begin{table}[hbp!]
\caption{\small Symmetry $\kk$-points of ORCF$_1$ and ORCF$_3$.}
\begin{tabular}{cccl|cccl}
\hline \hline
$\times \bb_1$ & $\times \bb_2$ & $\times \bb_3$ & & $\times \bb_1$ & $\times \bb_2$ & $\times \bb_3$ &\\
\hline
0 &  0		& 0 		&  $\Gamma$        &   0 &  $\eta$      & $\eta$       &  X\\
$\half$ &  $\half$+$\zeta$ & $\zeta$      &  A     &   1 &  1-$\eta$  & 1-$\eta$   &  X$_1$\\
$\half$ &  $\half$-$\zeta$ & 1-$\zeta$    &  A$_1$ &  $\half$ &  0    & $\half$     &  Y\\
$\half$ &  $\half$    & $\half$     &  L           &   $\half$ &  $\half$    & 0     &  Z\\
1 &  $\half$    & $\half$     &  T&    &     &      &  \\
\vspace{-3mm}\\
\multicolumn{4}{l}{$\zeta = (1+a^2/b^2-a^2/c^2)/4$}, & \multicolumn{4}{l}{$\eta = (1+a^2/b^2+a^2/c^2)/4$}\\
\hline
\end{tabular}
\end{table}
\end{center}

\vspace{-10mm}
\begin{figure}[hbp!]
\centerline{\epsfig{file=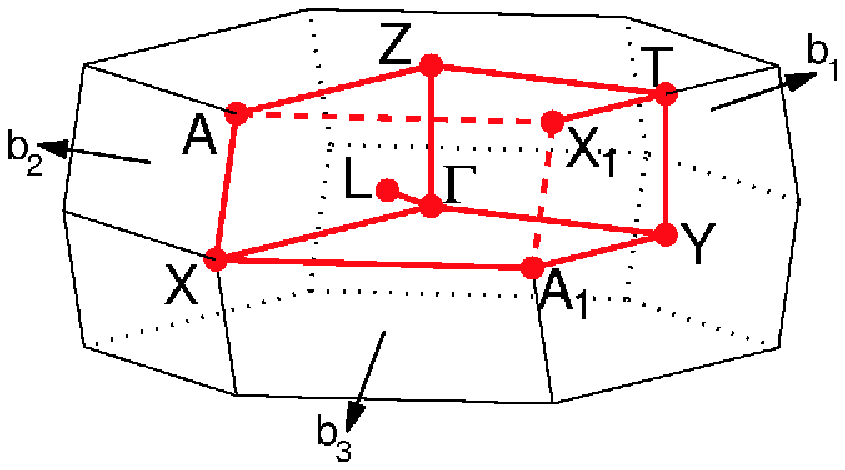,width=55mm}}
\vspace{-1mm}
\caption{\small 
Brillouin zone of ORCF$_1$ lattice. Path: $\Gamma$-Y-T-Z-$\Gamma$-X-A$_1$-Y$|$T-X$_1|$X-A-Z$|$L-$\Gamma$. An example of band structure using this path is given in Figure \ref{figbandORCF1}.}
\label{figORCF1}
\end{figure}

\vspace{-10mm}
\begin{figure}[hbp!]
\centerline{\epsfig{file=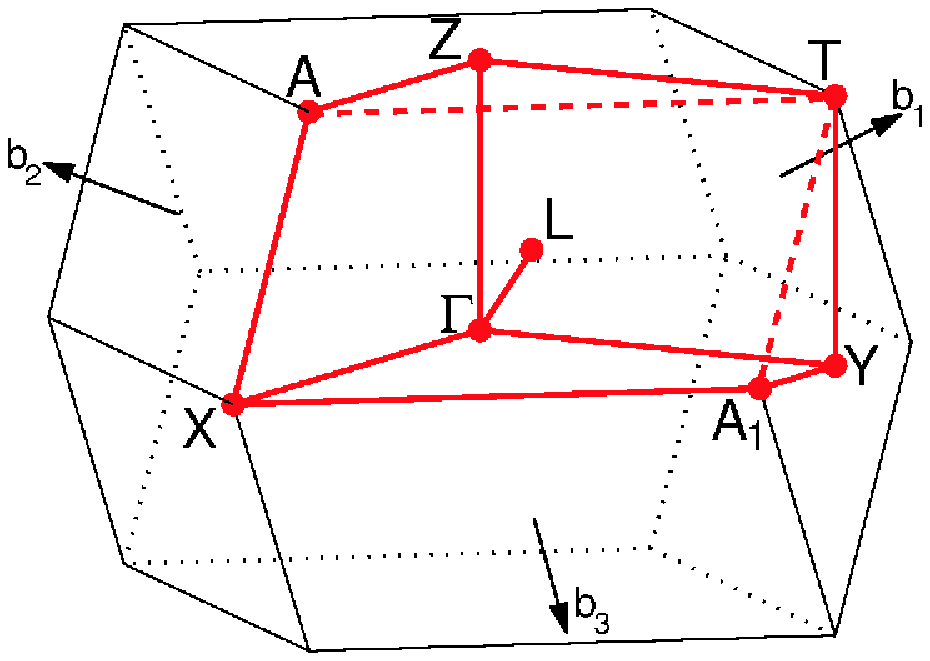,width=55mm}}
\vspace{-1mm}
\caption{\small 
Brillouin zone of ORCF$_3$ lattice. Path: $\Gamma$-Y-T-Z-$\Gamma$-X-A$_1$-Y$|$X-A-Z$|$L-$\Gamma$. An example of band structure using this path is given in Figure \ref{figbandORCF3}.}
\label{figORCF3}
\end{figure}

\vspace{-10mm}
\begin{center}
\begin{table}[hbp!]
\caption{\small Symmetry $\kk$-points of ORCF$_2$.}
\begin{tabular}{cccl|cccl}
\hline \hline
$\times \bb_1$ & $\times \bb_2$ & $\times \bb_3$ & & $\times \bb_1$ & $\times \bb_2$ & $\times \bb_3$ &\\
\hline 
0	& 0		& 0 		  & $\Gamma$     &1-$\phi$  & $\half$-$\phi$  	& $\half$ & H\\
$\half$     	 & $\half$-$\eta$ & 1-$\eta$    & C      &$\phi$    & $\half$+$\phi$  	& $\half$ & H$_1$\\
$\half$          & $\half$+$\eta$ & $\eta$      & C$_1$  & 0& $\half$ & $\half$ & X\\
$\half$-$\delta$ & $\half$    	  & 1-$\delta$  & D      & $\half$     	& 0  & $\half$ & Y\\
$\half$+$\delta$ & $\half$    	  & $\delta$    & D$_1$  & $\half$ & $\half$ & 0 & Z\\
$\half$     	& $\half$    	  & $\half$     & L & & & & \\
\vspace{-3mm}\\
\multicolumn{4}{l}{$\eta = (1+a^2/b^2-a^2/c^2)/4$}, & \multicolumn{4}{l}{$\delta = (1+b^2/a^2-b^2/c^2)/4$}\\
\multicolumn{4}{l}{$\phi = (1+c^2/b^2-c^2/a^2)/4$}\\
\hline
\end{tabular}
\end{table}
\end{center}

\vspace{-10mm}
\begin{figure}[hbp!]
\centerline{\epsfig{file=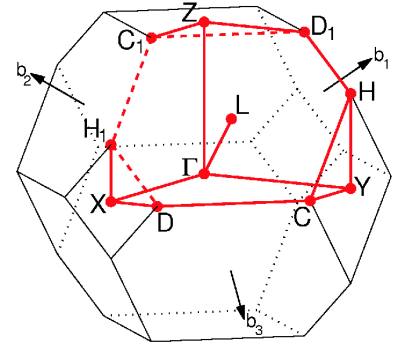,width=50mm}}
\vspace{-1mm}
\caption{\small 
Brillouin zone of ORCF$_2$ lattice. Path: $\Gamma$-Y-C-D-X-$\Gamma$-Z-D$_1$-H-C$|$C$_1$-Z$|$X-H$_1|$H-Y$|$L-$\Gamma$. An example of band structure using this path is given in Figure \ref{figbandORCF2}.}
\label{figORCF2}
\end{figure}

\pagebreak
\subsection{Body-centered Orthorhombic (ORCI, oI)}
\noindent
Ordering of the conventional lattice: $a<b<c$. \\
\begin{tabular}{l|l}
Conventional lattice: & Primitive lattice:\\
$\aa_1 = (a, 0, 0)$ & $\aa_1 = (-a/2,  b/2,  c/2)$\\
$\aa_2 = (0, b, 0)$ & $\aa_2 = ( a/2, -b/2,  c/2)$\\
$\aa_3 = (0, 0, c)$ & $\aa_3 = ( a/2,  b/2, -c/2)$\\
\end{tabular}

\begin{center}
\begin{table}[hbp!]
\caption{Symmetry $\kk$-points of ORCI.}
\begin{tabular}{cccl|cccl}
\hline \hline
$\times \bb_1$ & $\times \bb_2$ & $\times \bb_3$ & & $\times \bb_1$ & $\times \bb_2$ & $\times \bb_3$ &\\
\hline
0 & 0 & 0 & $\Gamma$    &  $\afth$ & $\afth$ & $\afth$ & W\\
-$\mu$ & $\mu$ & $\half$-$\delta$ & L        & -$\zeta$ & $\zeta$ & $\zeta$ & X\\
$\mu$ & -$\mu$ & $\half$+$\delta$ & L$_1$    &  $\zeta$ & 1-$\zeta$ & -$\zeta$ & X$_1$ \\
$\half$-$\delta$ & $\half$+$\delta$ & -$\mu$ & L$_2$     & $\eta$ & -$\eta$ & $\eta$ & Y\\
0 & $\half$ & 0 & R    & 1-$\eta$ & $\eta$ & -$\eta$ & Y$_1$\\
$\half$ & 0 & 0 & S    & $\half$ & $\half$ & -$\half$ & Z\\
0 & 0 & $\half$ & T    & & & &\\
\vspace{-3mm}\\
\multicolumn{4}{l}{$\zeta = (1+a^2/c^2)/4$}, & \multicolumn{4}{l}{$\delta = (b^2-a^2)/(4c^2)$}\\
\multicolumn{4}{l}{$\eta = (1+b^2/c^2)/4$}, & \multicolumn{4}{l}{$\mu = (a^2+b^2)/(4c^2)$}\\
\hline
\end{tabular}
\end{table}
\end{center}

\begin{figure}[hbp!]
\centerline{\epsfig{file=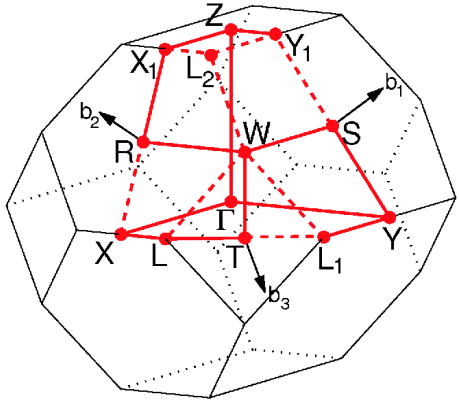,width=60mm}}
\vspace{-1mm}
\caption{\small 
Brillouin zone of ORCI lattice. Path: $\Gamma$-X-L-T-W-R-X$_1$-Z-$\Gamma$-Y-S-W$|$L$_1$-Y$|$Y$_1$-Z. An example of band structure using this path is given in Figure \ref{figbandORCI}.}
\label{figORCI}
\end{figure}

\subsection{C-centered Orthorhombic (ORCC, oS)}
\noindent
Ordering of the conventional lattice: $a<b$. \\
\begin{tabular}{l|l}
Conventional lattice: & Primitive lattice:\\
$\aa_1 = (a, 0, 0)$ & $\aa_1 = (a/2, -b/2, 0)$\\
$\aa_2 = (0, b, 0)$ & $\aa_2 = (a/2,  b/2, 0)$\\
$\aa_3 = (0, 0, c)$ & $\aa_3 = (0  ,  0  , c)$\\
\end{tabular}

\begin{center}
\begin{table}[h!]
\caption{Symmetry $\kk$-points of ORCC.}
\begin{tabular}{cccl|cccl}
\hline \hline
$\times \bb_1$ & $\times \bb_2$ & $\times \bb_3$ & & $\times \bb_1$ & $\times \bb_2$ & $\times \bb_3$ &\\
\hline
   0 &  0  & 0 &  $\Gamma$    &   -$\half$ &  $\half$  & $\half$ &  T\\
   $\zeta$  &  $\zeta$   & $\half$ &  A&  $\zeta$  &  $\zeta$   & 0 &  X\\
  -$\zeta$  &  1-$\zeta$ & $\half$ &  A$_1$&  -$\zeta$  &  1-$\zeta$ & 0 &  X$_1$ \\
  0 &  $\half$  & $\half$ &  R &  -$\half$ &  $\half$  & 0 &  Y\\
  0 &  $\half$  & 0 &  S &   0 &  0  & $\half$ &  Z\\
\vspace{-3mm}\\
\multicolumn{8}{l}{$\zeta=(1+a^2/b^2)/4$}\\
\hline
\end{tabular}
\end{table}
\end{center}

\begin{figure}[h!]
\centerline{\epsfig{file=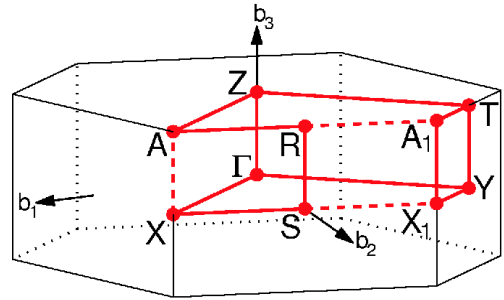,width=65mm}}
\vspace{-1mm}
\caption{\small 
Brillouin zone of ORCC lattice. Path: $\Gamma$-X-S-R-A-Z-$\Gamma$-Y-X$_1$-A$_1$-T-Y$|$Z-T. An example of band structure using this path is given in Figure \ref{figbandORCC}.}
\label{figORCC}
\end{figure}

\subsection{Hexagonal (HEX, hP)}
\noindent
Lattice:\\
$\aa_1 = (a/2, -(a\sqrt{3})/2, 0)$\\
$\aa_2 = (a/2,  (a\sqrt{3})/2, 0)$\\
$\aa_3 = (      0,                0, c)$\\

\vspace{-10mm}
\begin{center}
\begin{table}[hbp!]
\caption{Symmetry $\kk$-points of HEX.}
\begin{tabular}{cccl|cccl}
\hline \hline
$\times \bb_1$ & $\times \bb_2$ & $\times \bb_3$ & & $\times \bb_1$ & $\times \bb_2$ & $\times \bb_3$ &\\
\hline
   0   & 0   & 0   & $\Gamma$&   $\atrd$ & $\atrd$ & 0   & K\\
   0   & 0   & $\half$ & A&   $\half$ & 0   & $\half$ & L\\
   $\atrd$ & $\atrd$ & $\half$ & H&   $\half$ & 0   & 0   & M\\
\hline
\end{tabular}
\end{table}
\end{center}

\vspace{-5mm}
\begin{figure}[hbp!]
\centerline{\epsfig{file=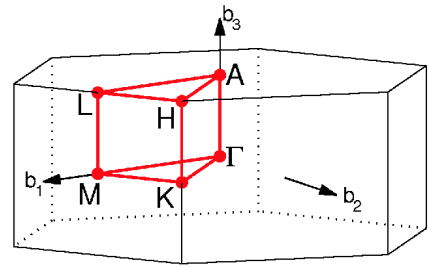,width=55mm}}
\vspace{-1mm}
\caption{\small 
Brillouin zone of HEX lattice. Path: $\Gamma$-M-K-$\Gamma$-A-L-H-A$|$L-M$|$K-H. An example of band structure using this path is given in Figure \ref{figbandHEX}.}
\label{figHEX}
\end{figure}

\pagebreak
\subsection{Rhombohedral (RHL, hR)}
\noindent
Lattice:\\
$\aa_1 = (a\cos(\alpha/2), -a\sin(\alpha/2), 0)$\\
$\aa_2 = (a\cos(\alpha/2),  a\sin(\alpha/2), 0)$\\
$\aa_3 = (a\cos\alpha/\cos(\alpha/2), 0, a\sqrt{1-\cos^2\alpha/\cos^2(\alpha/2)} \hspace*{1mm})$\\
Variations:\\
RHL$_1$: $\alpha<90^\circ$\\
RHL$_2$: $\alpha>90^\circ$\\

\vspace{-5mm}
\begin{center}
\begin{table}[hbp!]
\caption{Symmetry $\kk$-points of RHL$_1$.}
\begin{tabular}{cccl|cccl}
\hline \hline 
$\times \bb_1$ & $\times \bb_2$ & $\times \bb_3$ & & $\times \bb_1$ & $\times \bb_2$ & $\times \bb_3$ &\\
\hline 
0       & 0        & 0 	      & $\Gamma$& $\eta$ 	& $\nu$    & $\nu$    & P\\
$\eta$  & $\half$  & 1-$\eta$ & B       & 1-$\nu$ & 1-$\nu$ & 1-$\eta$  & P$_1$\\
$\half$ & 1-$\eta$ & $\eta$-1 & B$_1$   & $\nu$   & $\nu$   & $\eta$-1  & P$_2$\\
$\half$ & $\half$  & 0        & F       & 1-$\nu$ & $\nu$   & 0     	& Q\\
$\half$ & 0        & 0        & L       & $\nu$   & 0       & -$\nu$  	& X\\
0    	  & 0       & -$\half$  & L$_1$ & $\half$ & $\half$ & $\half$ 	& Z\\
\vspace{-3mm}\\
\multicolumn{8}{l}{$\eta = (1+4\cos\alpha)/(2+4\cos\alpha)$}\\
\multicolumn{8}{l}{$\nu = \tfth-\eta/2$}\\
\hline
\end{tabular}
\end{table}
\end{center}

\vspace{-5mm}
\begin{figure}[hbp!]
\centerline{\epsfig{file=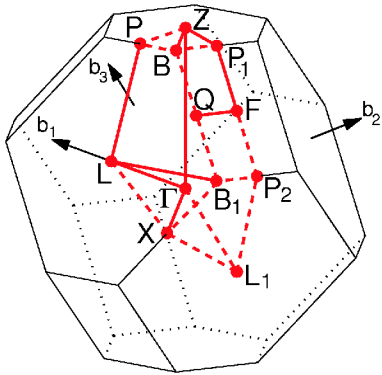,width=50mm}}
\vspace{-1mm}
\caption{\small 
Brillouin zone of RHL$_1$ lattice. Path: $\Gamma$-L-B$_1|$B-Z-$\Gamma$-X$|$Q-F-P$_1$-Z$|$L-P. An example of band structure using this path is given in Figure \ref{figbandRHL1}.}
\label{figRHL1}
\end{figure}

\vspace{-5mm}
\begin{center}
\begin{table}[hbp!]
\caption{Symmetry $\kk$-points of RHL$_2$.}
\begin{tabular}{cccl|cccl}
\hline \hline 
$\times \bb_1$ & $\times \bb_2$ & $\times \bb_3$ & & $\times \bb_1$ & $\times \bb_2$ & $\times \bb_3$ &\\
\hline 
0   &   0      &   0 &  $\Gamma$ & $\nu$    &  $\nu$-1 &  $\nu$-1 & P$_1$\\
$\half$ & -$\half$ & 0       & F & $\eta$   &  $\eta$  &  $\eta$  & Q\\
$\half$ & 0        & 0       & L & 1-$\eta$ & -$\eta$  & -$\eta$  & Q$_1$\\
1-$\nu$ & -$\nu$   & 1-$\nu$ & P & $\half$ & -$\half$  & $\half$  & Z\\
\vspace{-3mm}\\
\multicolumn{4}{l}{$\eta = 1/(2\tan^2(\alpha/2))$}, & \multicolumn{4}{l}{$\nu = \tfth-\eta/2$}\\
\hline
\end{tabular}
\end{table}
\end{center}

\vspace{-5mm}
\begin{figure}[htp!]
\centerline{\epsfig{file=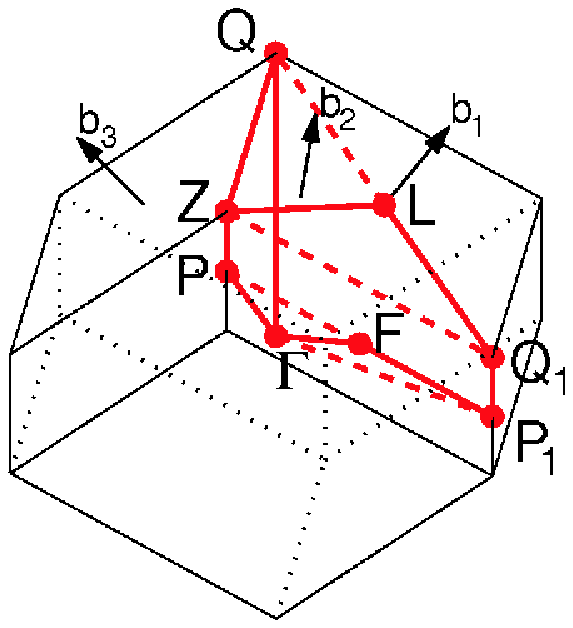,width=45mm}}
\vspace{-1mm}
\caption{\small
Brillouin zone of RHL$_2$ lattice. Path: $\Gamma$-P-Z-Q-$\Gamma$-F-P$_1$-Q$_1$-L-Z. An example of band structure using this path is given in Figure \ref{figbandRHL2}.}
\label{figRHL2}
\end{figure}

\subsection{Monoclinic (MCL, mP)}
\noindent
Ordering of the lattice: $a, b\leq c, \alpha<90^\circ, \beta=\gamma=90^\circ$.
Lattice:\\
$\aa_1 = (a, 0, 0)$\\
$\aa_2 = (0, b, 0)$\\
$\aa_3 = (0, c\cos\alpha, c\sin\alpha)$\\

\vspace{-10mm}
\begin{center}
\begin{table}[hbp!]
\caption{\small Symmetry $\kk$-points of MCL.}
\begin{tabular}{cccl|cccl}
\hline \hline \vspace{-3mm}\\
$\times \bb_1$ & $\times \bb_2$ & $\times \bb_3$ & & $\times \bb_1$ & $\times \bb_2$ & $\times \bb_3$ &\\
\hline \vspace{-3mm}\\
0  & 0  &  0  & $\Gamma$       & 0&   $\eta$  & -$\nu$     & H$_2$\\
$\half$& $\half$&  0  & A      & $\half$& $\eta$  & 1-$\nu$& M  \\
0  & $\half$&  $\half$& C      & $\half$& 1-$\eta$& $\nu$  & M$_1$\\
$\half$& 0  &  $\half$& D      & $\half$& $\eta$  & -$\nu$ & M$_2$  \\
$\half$& 0  & -$\half$& D$_1$  & 0&   $\half$     & 0      & X      \\
$\half$& $\half$&  $\half$& E  & 0&   0       & $\half$    & Y  \\
0  & $\eta$& 1-$\nu$& H        & 0 &  0 & -$\half$         & Y$_1$     \\
0&   1-$\eta$& $\nu$  & H$_1$  & $\half$& 0  &  0          & Z \\
 \vspace{-3mm}\\
\multicolumn{8}{l}{$\eta = (1-b\cos\alpha/c)/(2\sin^2\alpha)$}\\
\multicolumn{8}{l}{$\nu = \half-\eta c \cos\alpha/b$}\\
\hline
\end{tabular}
\end{table}
\end{center}

\vspace{-10mm}
\begin{figure}[hbp!]
\centerline{\epsfig{file=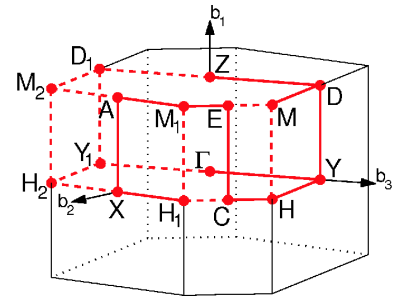,width=50mm}}
\vspace{-1mm}
\caption{\small 
Brillouin zone of MCL lattice. Path: $\Gamma$-Y-H-C-E-M$_1$-A-X-H$_1|$M-D-Z$|$Y-D. An example of band structure using this path is given in Figure \ref{figbandMCL}.}
\label{figMCL}
\end{figure}

\pagebreak
\subsection{C-centered Monoclinic (MCLC, mS)}
\noindent
Ordering of the conventional lattice: $a,b\leq c,\alpha<90^\circ,\beta=\gamma=90^\circ$.\\
\begin{tabular}{l|l}
Conventional lattice: & Primitive lattice:\\
$\aa_1 = (a, 0, 0)$ & $\aa_1 = (a/2, b/2, 0)$\\
$\aa_2 = (0, b, 0)$ & $\aa_2 = (-a/2, b/2, 0)$\\
$\aa_3 = (0, c\cos\alpha, c\sin\alpha)$ & $\aa_3 = (0, c\cos\alpha, c\sin\alpha)$\\
\end{tabular}

\noindent
Variations:\\
MCLC$_1$: $k_\gamma > 90^\circ$\\
MCLC$_2$: $k_\gamma = 90^\circ$\\
MCLC$_3$: $k_\gamma < 90^\circ,b\cos\alpha/c +b^2\sin^2\alpha/a^2 < 1$\\
MCLC$_4$: $k_\gamma < 90^\circ,b\cos\alpha/c +b^2\sin^2\alpha/a^2 = 1$\\
MCLC$_5$: $k_\gamma < 90^\circ,b\cos\alpha/c +b^2\sin^2\alpha/a^2 > 1$\\

\vspace{-10mm}
\begin{center}
\begin{table}[hbp!]
\caption{\small Symmetry $\kk$-points of MCLC$_1$ and MCLC$_2$.}
\begin{tabular}{cccl|cccl}
\hline \hline \vspace{-3mm}\\
$\times \bb_1$ & $\times \bb_2$ & $\times \bb_3$ & & $\times \bb_1$ & $\times \bb_2$ & $\times \bb_3$ &\\
\hline \vspace{-3mm}\\
0   & 0  & 0 		            & $\Gamma$ & $\half$  & $\half$  & $\half$ & L\\
$\half$  & 0 		 & 0  	       & N     & $\half$  & 0 	     & $\half$ & M\\
0 	 & -$\half$  	 & 0  	       & N$_1$ & 1-$\psi$ & $\psi$-1 & 0       & X\\
1-$\zeta$  & 1-$\zeta$& 1-$\eta$       & F     & $\psi$   & 1-$\psi$ & 0       & X$_1$\\
$\zeta$ & $\zeta$ & $\eta$             & F$_1$ & $\psi$-1 & -$\psi$  & 0       & X$_2$\\
-$\zeta$   & -$\zeta$ & 1-$\eta$       & F$_2$ & $\half$  & $\half$  & 0       & Y\\
1-$\zeta$ & -$\zeta$ & 1-$\eta$        & F$_3$ & -$\half$ & -$\half$ & 0       & Y$_1$\\
$\phi$  	  & 1-$\phi$ & $\half$ & I     & 0  	  & 0 	     & $\half$ & Z\\
1-$\phi$& $\phi$-1& $\half$ 	       & I$_1$ & & & & \\
\vspace{-3mm}\\
\multicolumn{8}{l}{$\zeta = (2 - b\cos\alpha/c)/(4\sin^2\alpha)$}\\
\multicolumn{8}{l}{$\eta = \half + 2\zeta c\cos\alpha/b$}\\
\multicolumn{8}{l}{$\psi = \tfth - a^2/(4b^2\sin^2\alpha)$}\\
\multicolumn{8}{l}{$\phi = \psi+(\tfth-\psi)b\cos\alpha/c$}\\
\hline
\end{tabular}
\end{table}
\end{center}

\vspace{-10mm}

\vspace{-10mm}
\begin{figure}[hbp!]
\centerline{\epsfig{file=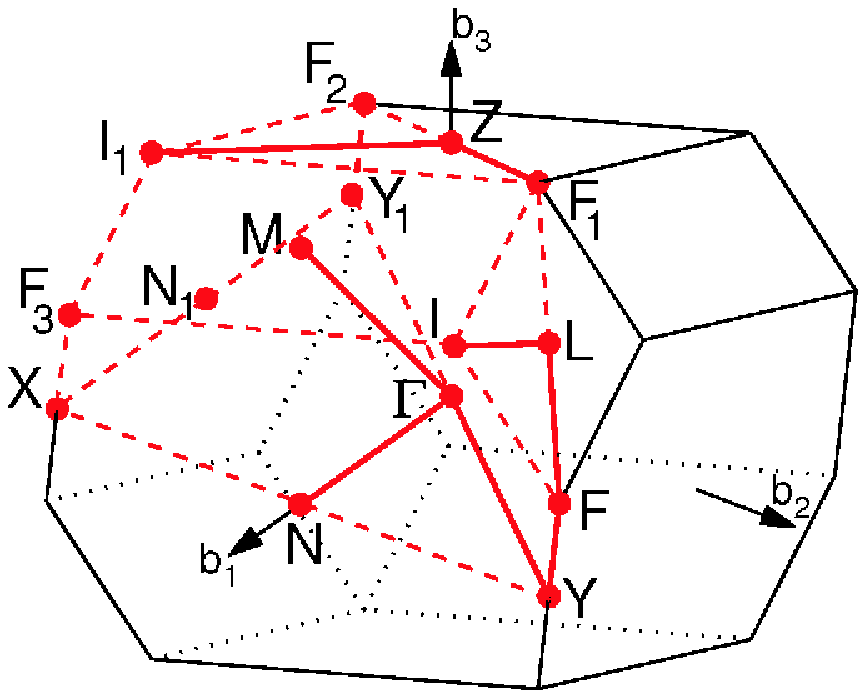
,width=60mm}}
\vspace{-1mm}
\caption{\small 
Brillouin zone of MCLC$_2$ lattice. Note that Y is equivalent to X. Path: $\Gamma$-Y-F-L-I$|$I$_1$-Z-F$_1|$N-$\Gamma$-M. An example of band structure using this path is given in Figure \ref{figbandMCLC2}.}
\label{figMCLC2}
\end{figure}

\vspace{-10mm}
\begin{center}
\begin{table}[hbp!]
\caption{\small Symmetry $\kk$-points of MCLC$_3$ and MCLC$_4$.}
\begin{tabular}{cccl|cccl}
\hline \hline 
$\times \bb_1$ & $\times \bb_2$ & $\times \bb_3$ & & $\times \bb_1$ & $\times \bb_2$ & $\times \bb_3$ &\\
\hline 
0 	&   0 	&   0 	         & $\Gamma$ & $\half$ 	&  0 	&  0   	 & N\\
1-$\phi$ & 1-$\phi$ & 1-$\psi$   & F        & 0 	& -$\half$ 	&  0   	 & N$_1$\\
$\phi$   &  $\phi$-1 &  $\psi$   & F$_1$    & $\half$ & -$\half$ 	&  0   	 & X\\
1-$\phi$ & -$\phi$ & 1-$\psi$    & F$_2$    & $\mu$    &  $\mu$    &  $\delta$ & Y\\
$\zeta$  &  $\zeta$  & $\eta$    & H        & 1-$\mu$  & -$\mu$    & -$\delta$ & Y$_1$\\
1-$\zeta$ & -$\zeta$ &  1-$\eta$ & H$_1$    & -$\mu$    & -$\mu$   & -$\delta$ & Y$_2$\\
-$\zeta$  & -$\zeta$  & 1-$\eta$ & H$_2$    & $\mu$    &  $\mu$-1  &  $\delta$ & Y$_3$\\
$\half$ & -$\half$ &  $\half$    & I        & 0 	&  0 	&  $\half$   	 & Z\\
$\half$ &  0 	&  $\half$       & M        & & & & \\
\vspace{-3mm}\\
\multicolumn{8}{l}{$\mu = (1 + b^2/a^2)/4$}\\
\multicolumn{8}{l}{$\delta = bc\cos\alpha/(2a^2)$}\\
\multicolumn{8}{l}{$\zeta = \mu-\afth+(1-b\cos\alpha/c)/(4\sin^2\alpha)$}\\
\multicolumn{8}{l}{$\eta = \half + 2\zeta c\cos\alpha/b$}\\
\multicolumn{8}{l}{$\phi = 1+\zeta-2\mu$}\\
\multicolumn{8}{l}{$\psi = \eta-2\delta$}\\
\hline
\end{tabular}
\end{table}
\end{center}

\vspace{-10mm}
\begin{figure}[hbp!]
\centerline{\epsfig{file=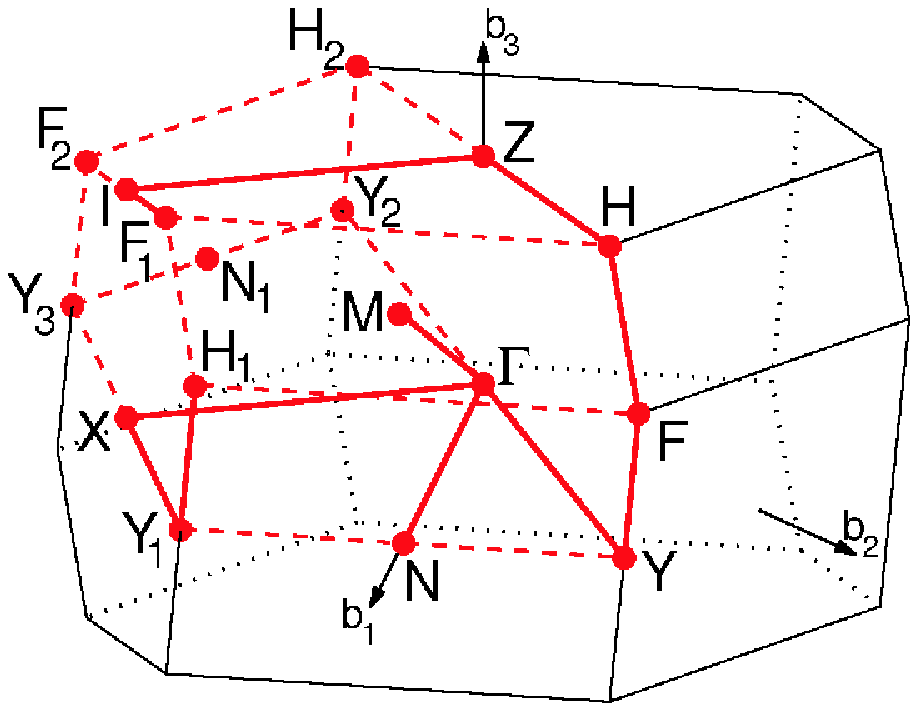
,width=60mm}}
\vspace{-1mm}
\caption{\small 
Brillouin zone of MCLC$_3$ lattice. Path: $\Gamma$-Y-F-H-Z-I-F$_1|$H$_1$-Y$_1$-X-$\Gamma$-N$|$M-$\Gamma$. An example of band structure using this path is given in Figure \ref{figbandMCLC3}.}
\label{figMCLC3}
\end{figure}

\vspace{-10mm}
\begin{figure}[hbp!]
\centerline{\epsfig{file=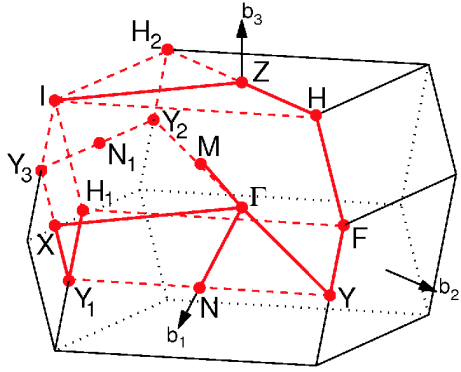
,width=60mm}}
\vspace{-1mm}
\caption{\small 
Brillouin zone of MCLC$_4$ lattice. Note that I is equivalent to F. Path: $\Gamma$-Y-F-H-Z-I$|$H$_1$-Y$_1$-X-$\Gamma$-N$|$M-$\Gamma$. An example of band structure using this path is given in Figure \ref{figbandMCLC4}.}
\label{figMCLC4}
\end{figure}

\vspace{-10mm}
\begin{center}
\begin{table}[hbp!]
\caption{\small Symmetry $\kk$-points of MCLC$_5$.}
\begin{tabular}{cccl|cccl}
\hline \hline 
$\times \bb_1$ & $\times \bb_2$ & $\times \bb_3$ & & $\times \bb_1$ & $\times \bb_2$ & $\times \bb_3$ &\\
\hline 
0 	&   	0 &   0 	& $\Gamma$ & $\half$ &    0   &  $\half$   	 & M\\
$\nu$   &  $\nu$  &  $\omega$    & F       & $\half$ & 	0  &   0         & N\\
1-$\nu$ &  1-$\nu$ &  1-$\omega$ & F$_1$   & 0 	&   -$\half$ &  0      	 & N$_1$\\
$\nu$   &  $\nu$-1 &  $\omega$   & F$_2$   & $\half$ & -$\half$ &  0        & X\\
$\zeta$  &  $\zeta$ &  $\eta$    & H       & $\mu$ &  $\mu$&  $\delta$        & Y\\
1-$\zeta$ & -$\zeta$ &   1-$\eta$ & H$_1$  & 1-$\mu$ &  -$\mu$  & -$\delta$   & Y$_1$\\
-$\zeta$   & -$\zeta$ &  1-$\eta$ & H$_2$  & -$\mu$ & -$\mu$ & -$\delta$      & Y$_2$\\
$\rho$   &  1-$\rho$ &  $\half$  & I       & $\mu$ &  $\mu$-1 & $\delta$      & Y$_3$\\
1-$\rho$ &  $\rho$-1 & $\half$   & I$_1$   & 0 & 0  & $\half$                 & Z\\
$\half$ & $\half$  &  $\half$    & L       & & & & \\
\vspace{-3mm}\\
\multicolumn{8}{l}{$\zeta = (b^2/a^2+(1-b\cos\alpha/c)/\sin^2\alpha)/4$}\\
\multicolumn{8}{l}{$\mu = \eta/2+b^2/(4a^2)-bc\cos\alpha/(2a^2)$}\\
\multicolumn{8}{l}{$\omega = (4\nu-1-b^2\sin^2\alpha/a^2)c/(2b\cos\alpha)$}\\
\multicolumn{5}{l}{$\eta = \half + 2\zeta c\cos\alpha/b$}, & \multicolumn{3}{l}{$\nu = 2\mu-\zeta$}\\
\multicolumn{5}{l}{$\delta = \zeta c\cos\alpha/b + \omega/2 - \afth$}, & \multicolumn{3}{l}{$\rho = 1-\zeta a^2/b^2$}\\
\hline
\end{tabular}
\end{table}
\end{center}

\vspace{-10mm}
\begin{figure}[hbp!]
\centerline{\epsfig{file=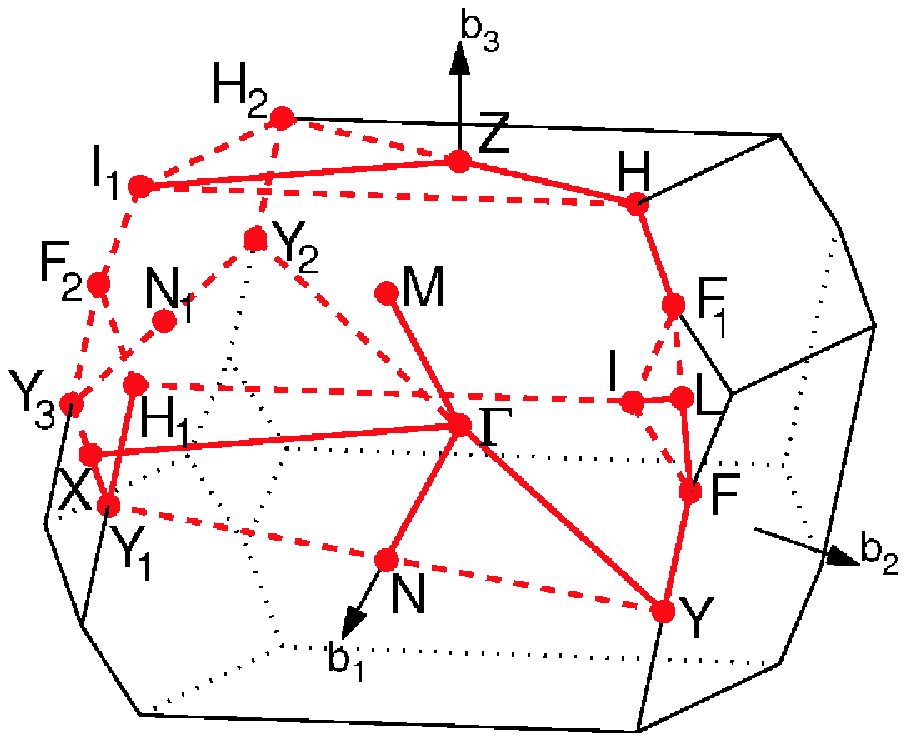
,width=60mm}}
\vspace{-1mm}
\caption{\small 
Brillouin zone of MCLC$_5$ lattice. Path: $\Gamma$-Y-F-L-I$|$I$_1$-Z-H-F$_1|$H$_1$-Y$_1$-X-$\Gamma$-N$|$M-$\Gamma$. An example of band structure using this path is given in Figure \ref{figbandMCLC5}.}
\label{figMCLC5}
\end{figure}

\pagebreak
\subsection{Triclinic (TRI, aP)}
\noindent
Lattice:\\
$\aa_1 = (a, 0, 0)$ \\
$\aa_2 = (b\cos\gamma, b\sin\gamma, 0)$ \\
$\aa_3 = (c\cos\beta,\frac{c}{\sin\gamma}[\cos\alpha -\cos\beta\cos\gamma],$\\
$\hspace*{6mm}\frac{c}{\sin\gamma}\sqrt{\sin^2\gamma-\cos^2\alpha-\cos^2\beta+2\cos\alpha \cos\beta \cos\gamma} \hspace{1mm})$ \\
Variations:\\
TRI$_{1a}$ : $k_\alpha>90^\circ , k_\beta>90^\circ , k_\gamma>90^\circ$\\
TRI$_{1b}$ : $k_\alpha<90^\circ , k_\beta<90^\circ , k_\gamma<90^\circ$\\
TRI$_{2a}$ : $k_\alpha>90^\circ , k_\beta>90^\circ , k_\gamma=90^\circ$\\
TRI$_{2b}$ : $k_\alpha<90^\circ , k_\beta<90^\circ , k_\gamma=90^\circ$\\

\vspace{-10mm}
\begin{center}
\begin{table}[hbp!]
\caption{\small Symmetry $\kk$-points of TRI$_{1a}$ and TRI$_{2a}$.}
\begin{tabular}{cccl|cccl}
\hline \hline \vspace{-3mm}\\
$\times \bb_1$ & $\times \bb_2$ & $\times \bb_3$ & & $\times \bb_1$ & $\times \bb_2$ & $\times \bb_3$ &\\
\hline \vspace{-3mm}\\
   0   &  0   & 0   & $\Gamma$ & $\half$ & $\half$ & $\half$ &	R\\
   $\half$ &  $\half$ & 0   & L        & $\half$ & 0   & 0   & X\\
   0   &  $\half$ & $\half$ & M        & 0   & $\half$ & 0   &	Y\\
   $\half$ &  0   & $\half$ & N        & 0   & 0   & $\half$ &	Z\\
\hline
\end{tabular}
\end{table}
\end{center}

\vspace{-15mm}
\begin{figure}[hbp!]
\centerline{\epsfig{file=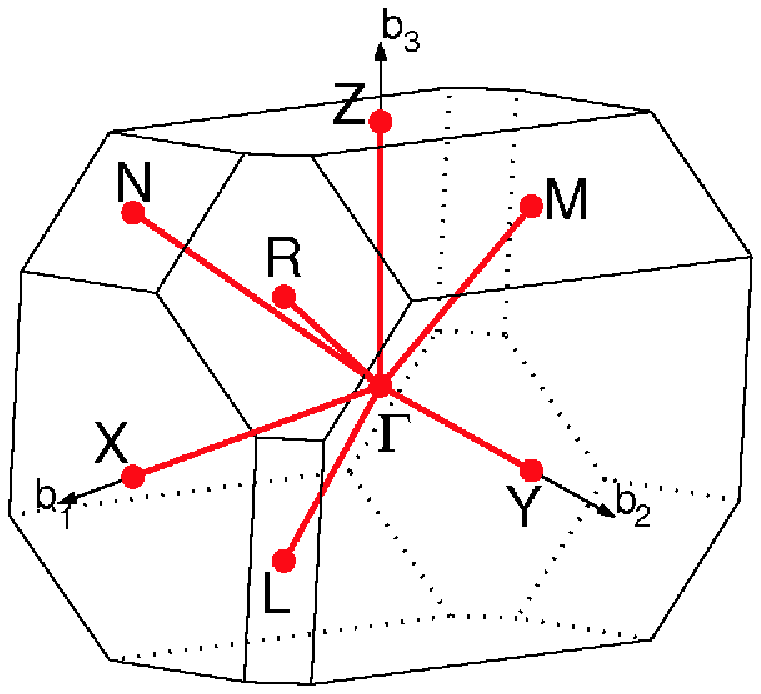
,width=55mm}}
\caption{\small 
Brillouin zone of TRI$_{1a}$ lattice. Path: X-$\Gamma$-Y$|$L-$\Gamma$-Z$|$N-$\Gamma$-M$|$R-$\Gamma$. An example of band structure using this path is given in Figure \ref{figbandTRI1a}.}
\label{figTRI1a}
\end{figure}

\vspace{-10mm}
\begin{figure}[hbp!]
\centerline{\epsfig{file=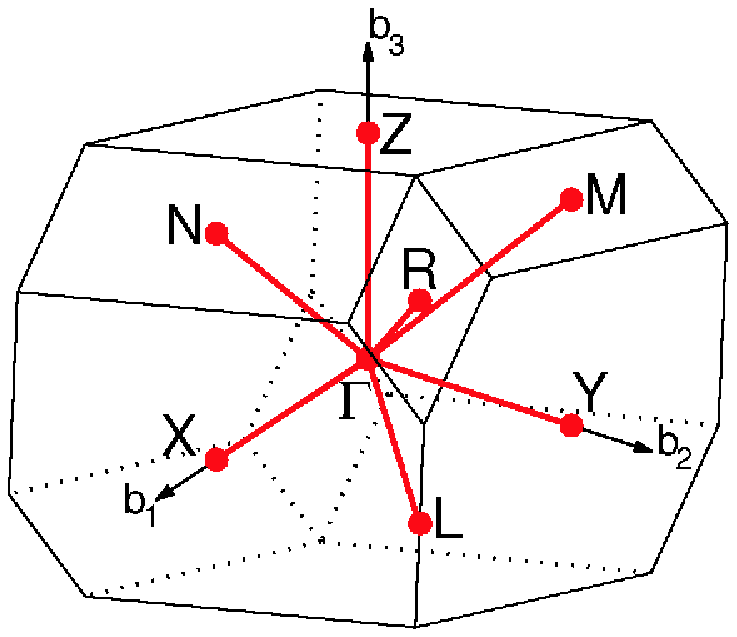
,width=55mm}}
\caption{\small
Brillouin zone of TRI$_{2a}$ lattice. Path: X-$\Gamma$-Y$|$L-$\Gamma$-Z$|$N-$\Gamma$-M$|$R-$\Gamma$.}
\label{figTRI2a}
\end{figure}

\vspace{-10mm}
\begin{center}
\begin{table}[hbp!]
\caption{\small Symmetry $\kk$-points of TRI$_{1b}$ and TRI$_{2b}$.}
\begin{tabular}{cccl|cccl}
\hline \hline \vspace{-3mm}\\
$\times \bb_1$ & $\times \bb_2$ & $\times \bb_3$ & & $\times \bb_1$ & $\times \bb_2$ & $\times \bb_3$ &\\
\hline \vspace{-3mm}\\
 0   &  0   & 0   & $\Gamma$ &  0   & -$\half$ & $\half$ & R\\
 $\half$ & -$\half$ & 0   & L        &  0   & -$\half$ & 0   & X\\
 0   &  0   & $\half$ & M        &  $\half$ &  0   & 0   & Y\\
-$\half$ & -$\half$ & $\half$ & N        & -$\half$ &  0   & $\half$ & Z\\
\hline
\end{tabular}
\end{table}
\end{center}

\vspace{-10mm}
\begin{figure}[hbp!]
\centerline{\epsfig{file=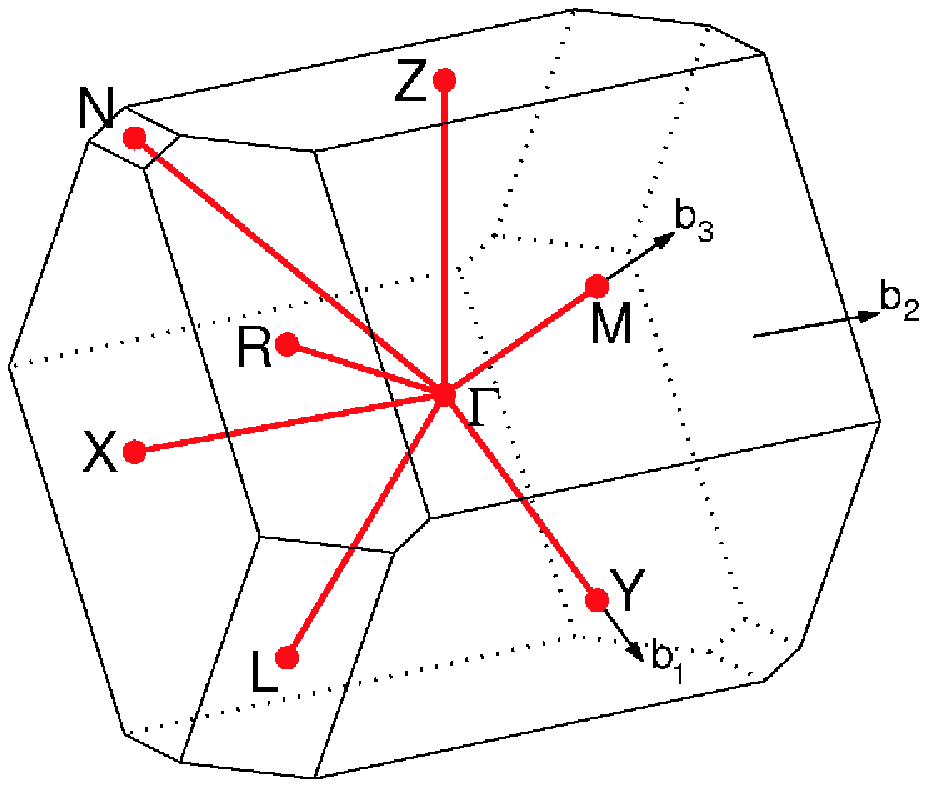
,width=60mm}}
\caption{\small
Brillouin zone of TRI$_{1b}$ lattice. Path: X-$\Gamma$-Y$|$L-$\Gamma$-Z$|$N-$\Gamma$-M$|$R-$\Gamma$. An example of band structure using this path is given in Figure \ref{figbandTRI1b}.}
\label{figTRI1b}
\end{figure}

\vspace{-10mm}
\begin{figure}[hbp!]
\centerline{\epsfig{file=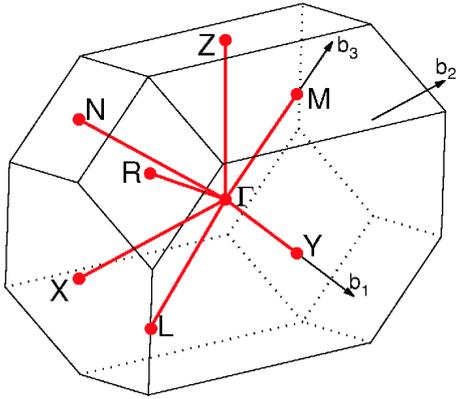
,width=60mm}}
\caption{\small
Brillouin zone of TRI$_{2b}$ lattice. Path: X-$\Gamma$-Y$|$L-$\Gamma$-Z$|$N-$\Gamma$-M$|$R-$\Gamma$.}
\label{figTRI2b}
\end{figure}

\pagebreak
\section{Appendix B}
\label{appendixB}
Selected examples of band structure for each shape of Brillouin zone are presented in this section.
The Fermi energy is shifted to the valence band maximum at zero.
In each figure, the orbital-projected total density of states $N(E)$ are plotted in the right panel in logarithmic scale.

\begin{figure}[h!]
\centerline{\epsfig{file=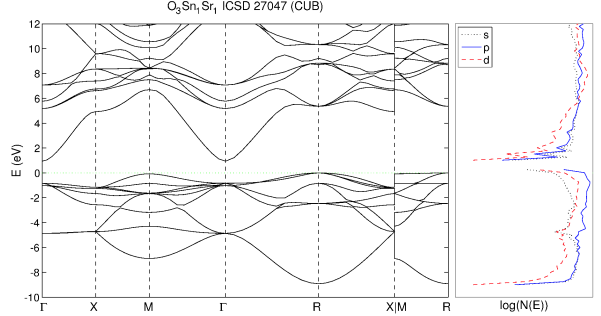
,height=42mm}}
\vspace{-3mm}
\caption{\small
Band structure of Sr(SnO$_3$) in CUB lattice.}
\label{figbandCUB}
\end{figure}
%
\begin{figure}[h!]
\centerline{\epsfig{file=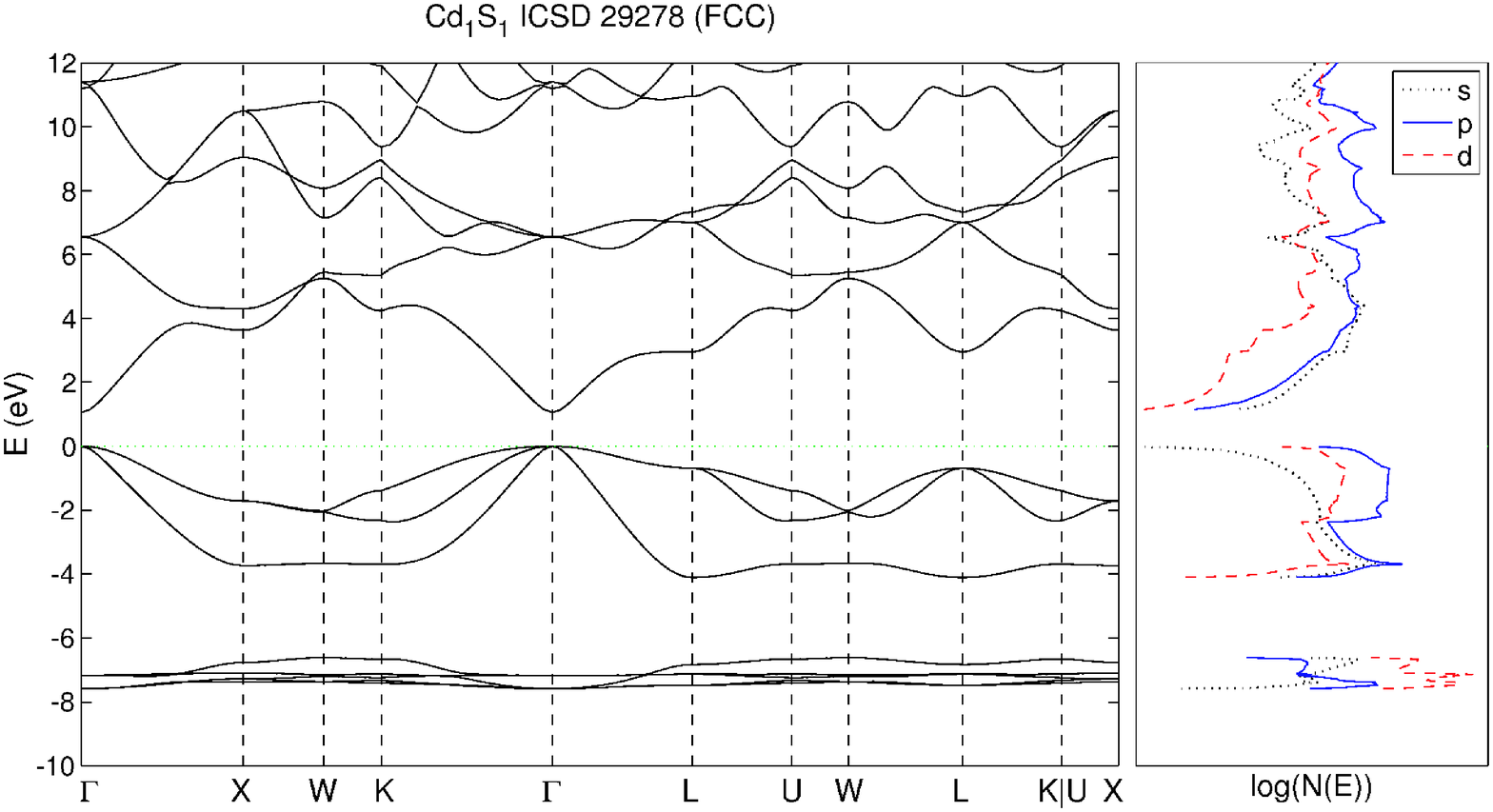
,height=42mm}}
\vspace{-3mm}
\caption{\small
Band structure of CdS in FCC lattice.}
\label{figbandFCC}
\end{figure}
%
\clearpage 
\begin{figure}[h!]
\centerline{\epsfig{file=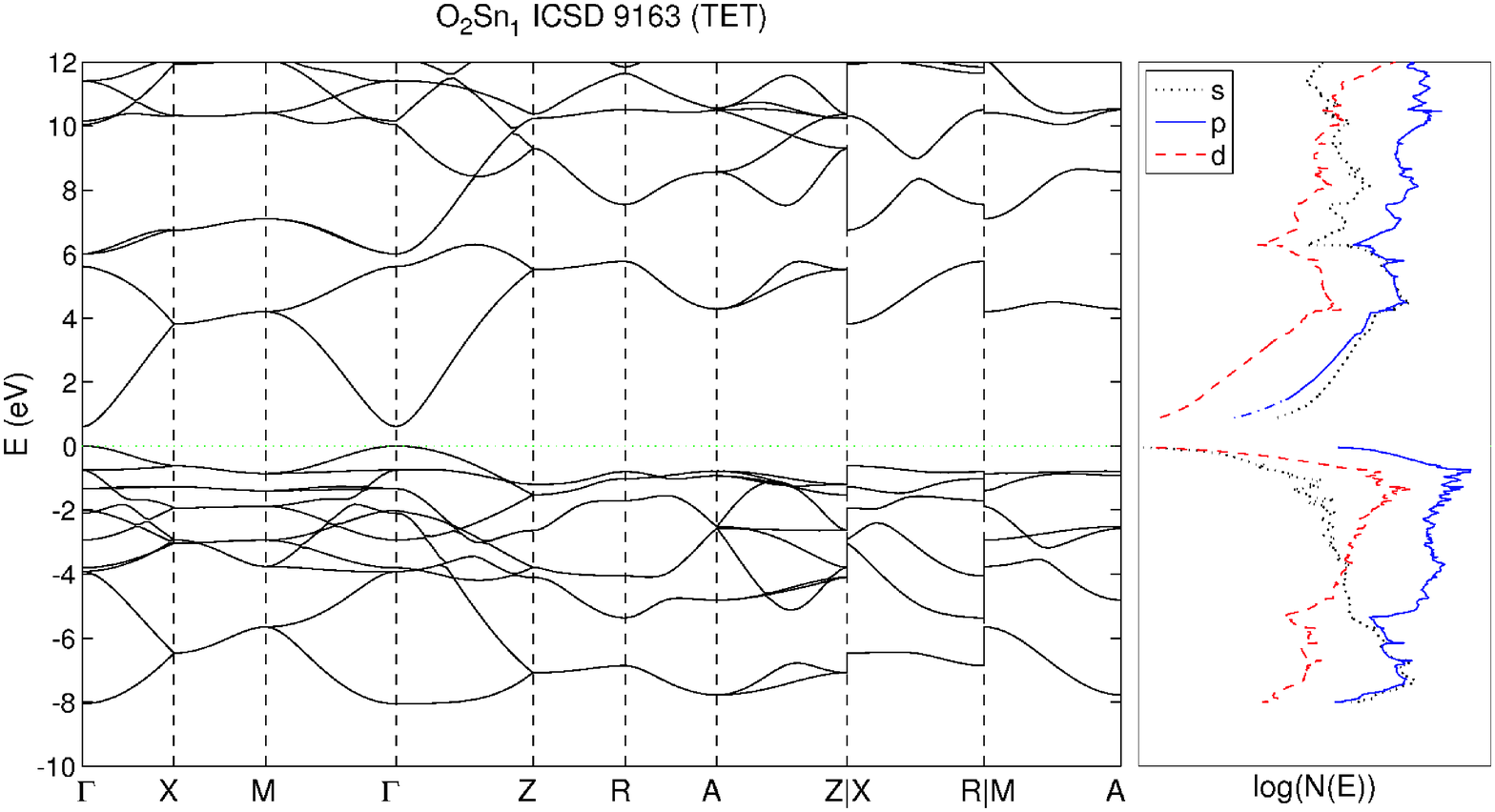
,height=42mm}}
\vspace{-3mm}
\caption{\small
Band structure of SnO$_2$ in TET lattice.}
\label{figbandTET}
\end{figure}
%
\begin{figure}[h!]
\centerline{\epsfig{file=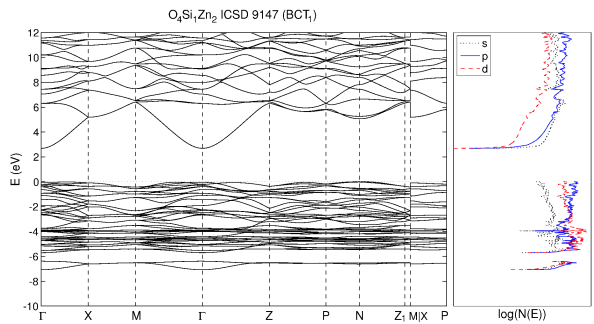
,height=42mm}}
\vspace{-3mm}
\caption{\small
Band structure of Zn$_2$(SiO$_4$) in BCT$_1$ lattice.}
\label{figbandBCT1}
\end{figure}
%
\begin{figure}[h!]
\centerline{\epsfig{file=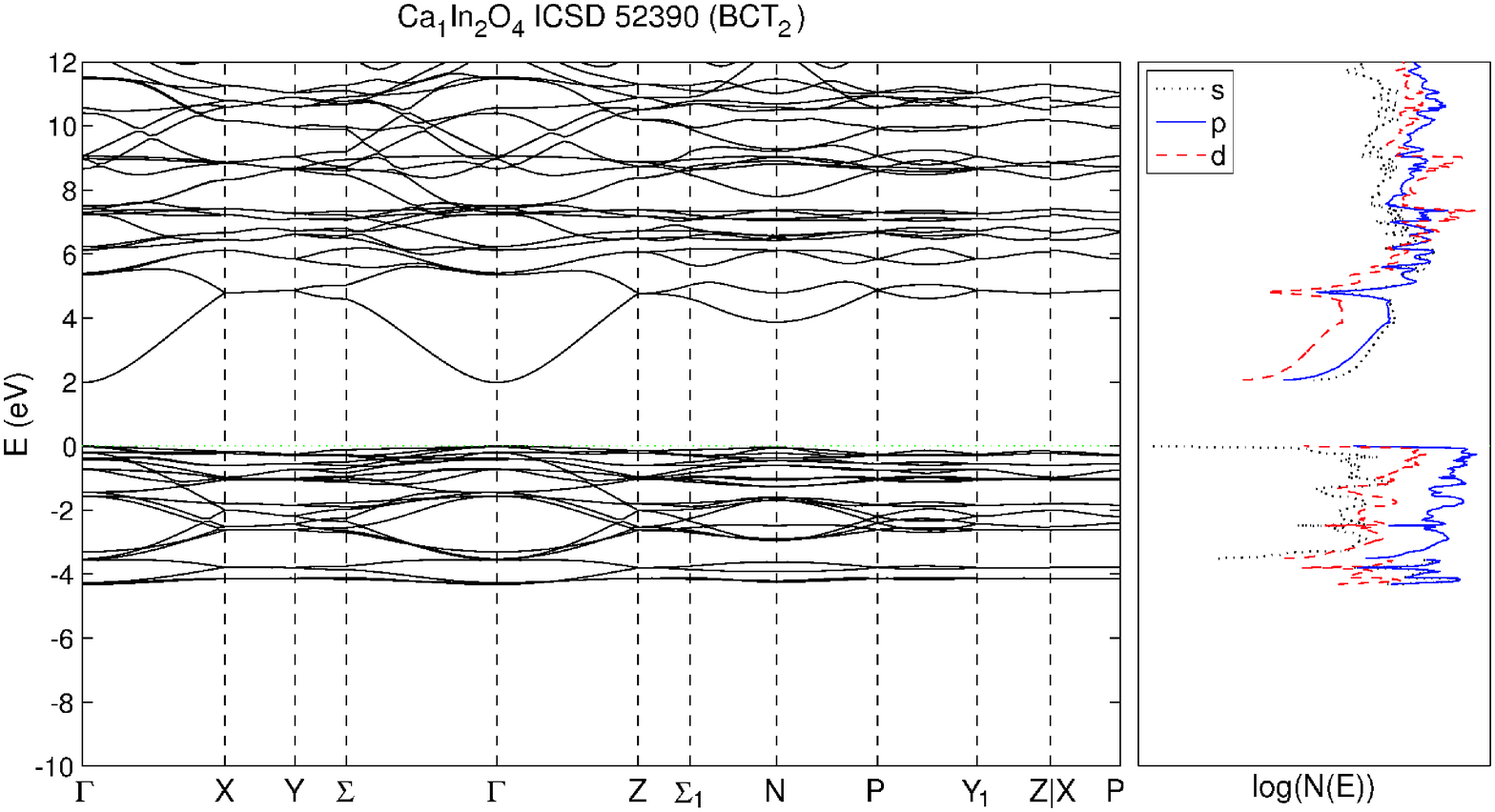
,height=42mm}}
\vspace{-3mm}
\caption{\small
Band structure of CaIn$_2$O$_4$ in BCT$_2$ lattice.}
\label{figbandBCT2}
\end{figure}
%
\begin{figure}[h!]
\centerline{\epsfig{file=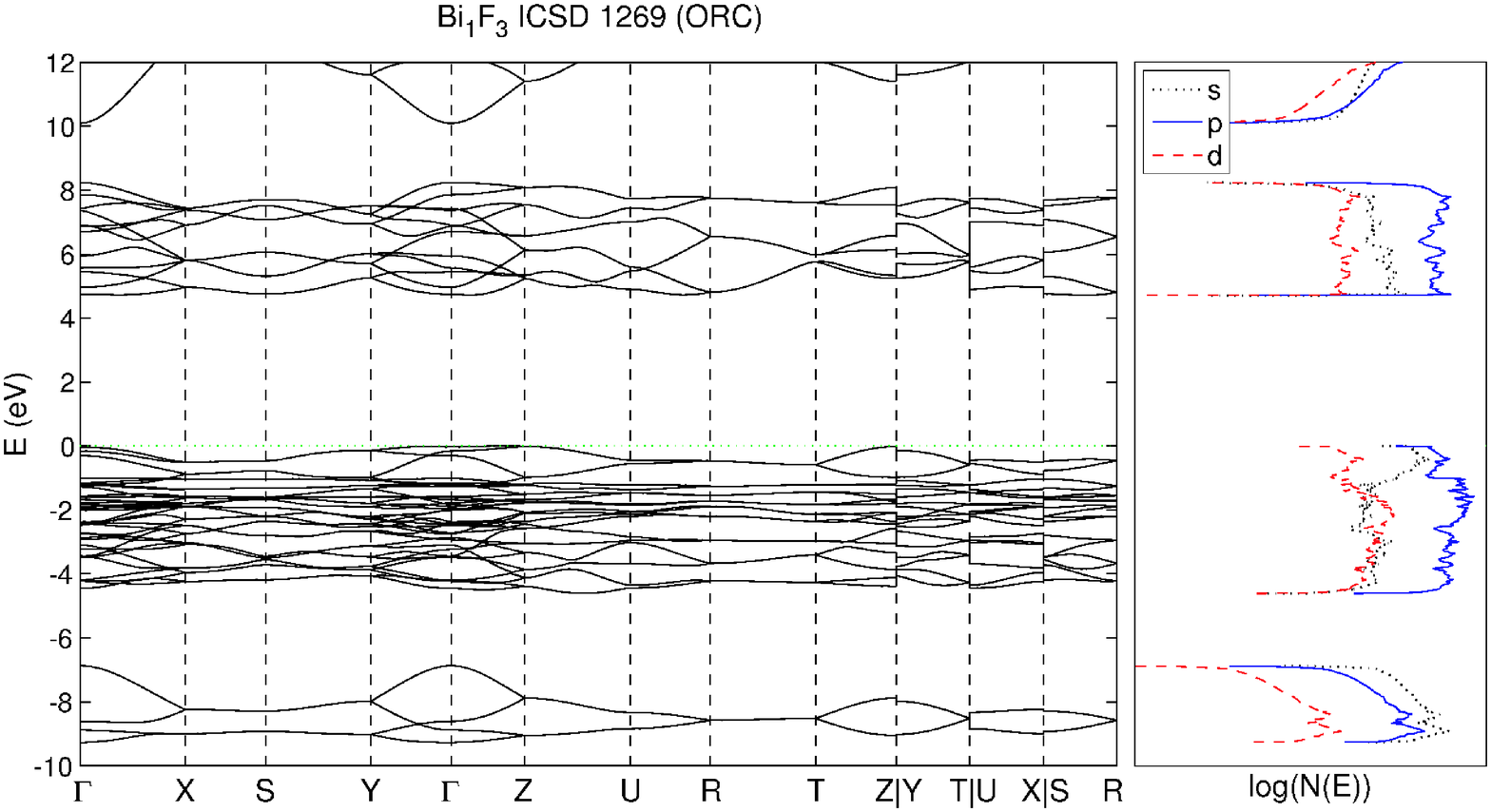
,height=42mm}}
\vspace{-3mm}
\caption{\small
Band structure of BiF$_3$ in ORC lattice.}
\label{figbandORC}
\end{figure}


\begin{figure}[h!]
\centerline{\epsfig{file=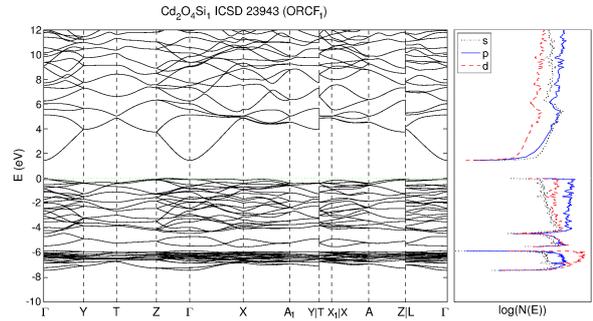
,height=42mm}}
\vspace{-3mm}
\caption{\small
Band structure of Cd$_2$(SiO$_4$) in ORCF$_1$ lattice.}
\label{figbandORCF1}
\end{figure}
%
%
\begin{figure}[h!]
\centerline{\epsfig{file=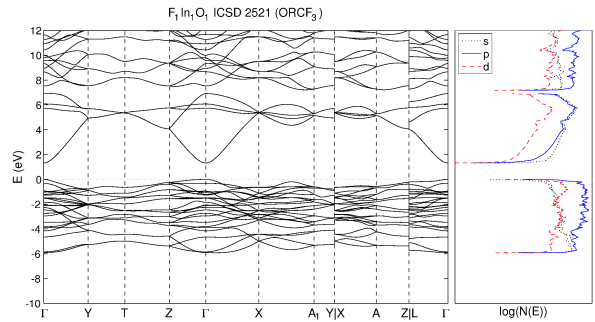
,height=42mm}}
\vspace{-3mm}
\caption{\small
Band structure of InOF in ORCF$_3$ lattice.}
\label{figbandORCF3}
\end{figure}
%
\begin{figure}[h!]
\centerline{\epsfig{file=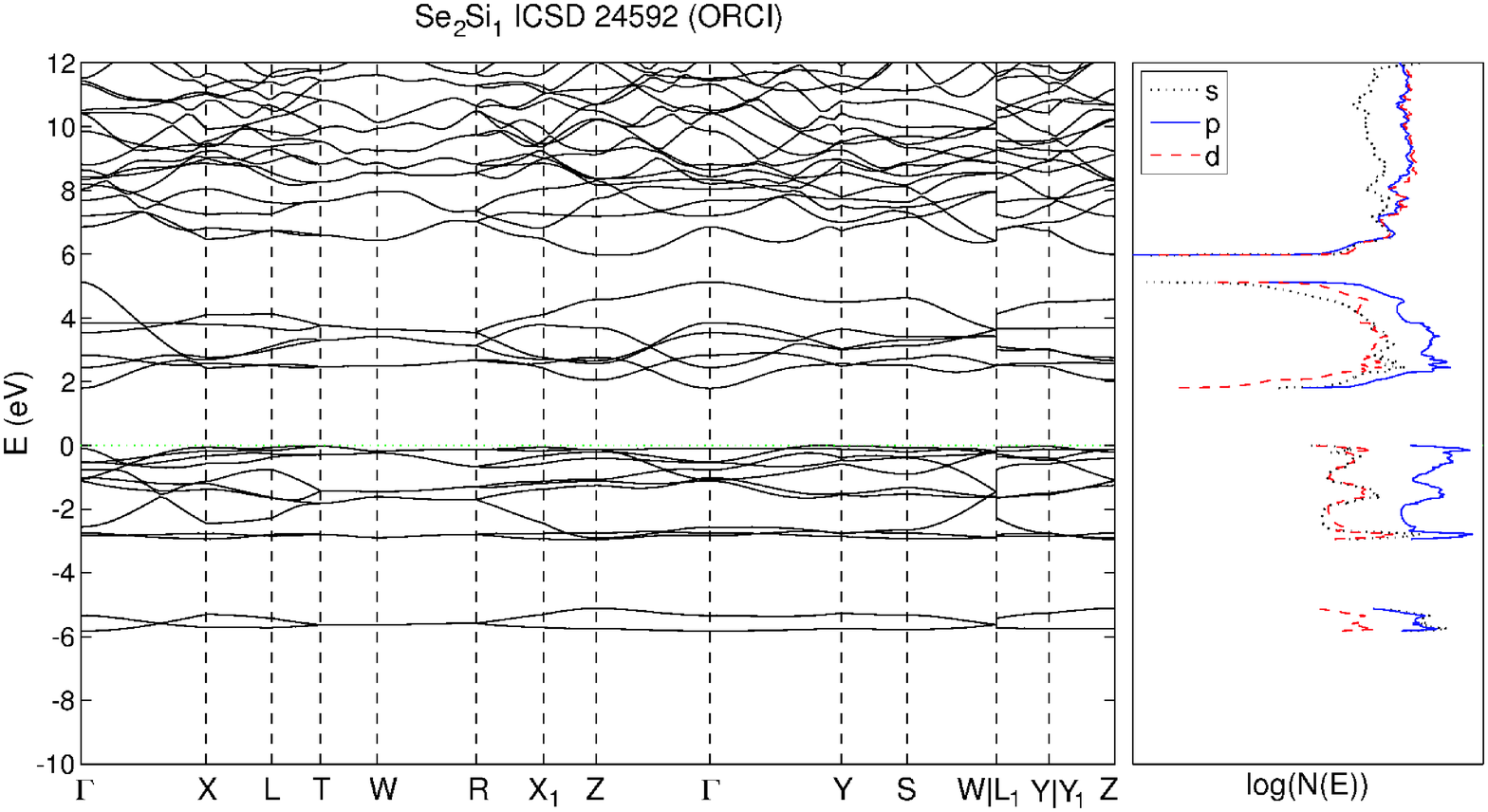
,height=42mm}}
\vspace{-3mm}
\caption{\small
Band structure of SiSe$_2$ in ORCI lattice.}
\label{figbandORCI}
\end{figure}
\clearpage
\begin{figure}[h!]
\centerline{\epsfig{file=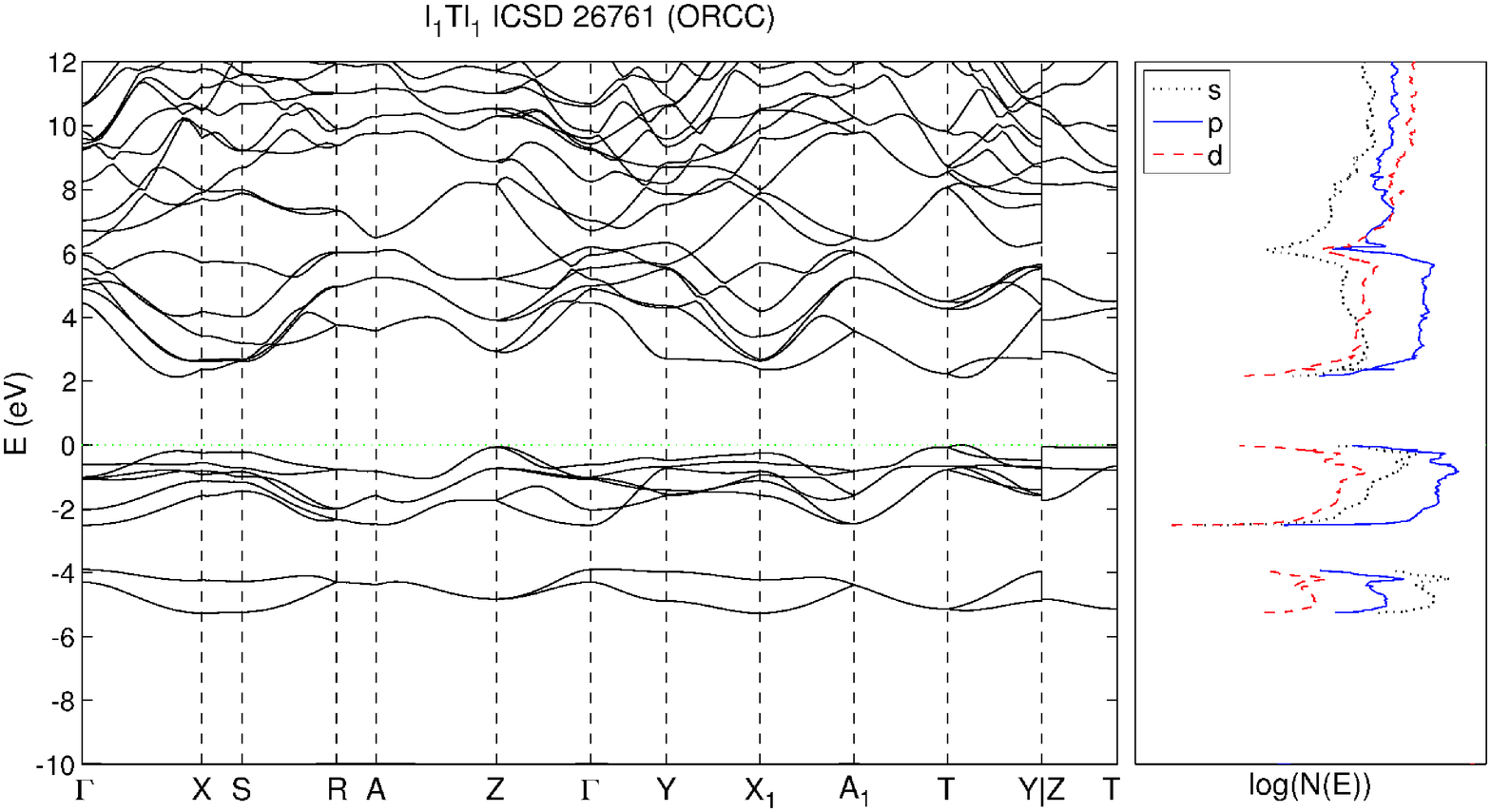
,height=42mm}}
\vspace{-3mm}
\caption{\small
Band structure of TlI in ORCC lattice.}
\label{figbandORCC}
\end{figure}
%
\begin{figure}[h!]
\centerline{\epsfig{file=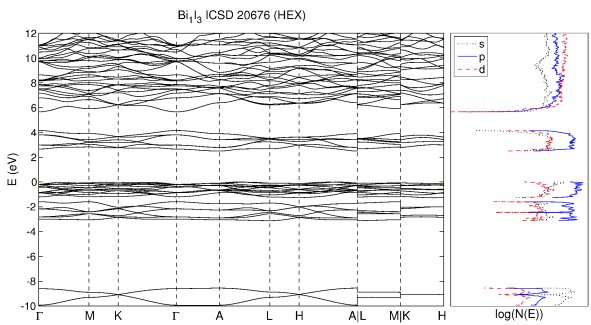
,height=42mm}}
\vspace{-3mm}
\caption{\small
Band structure of BiI$_3$ in HEX lattice.}
\label{figbandHEX}
\end{figure}
%
\begin{figure}[h!]
\centerline{\epsfig{file=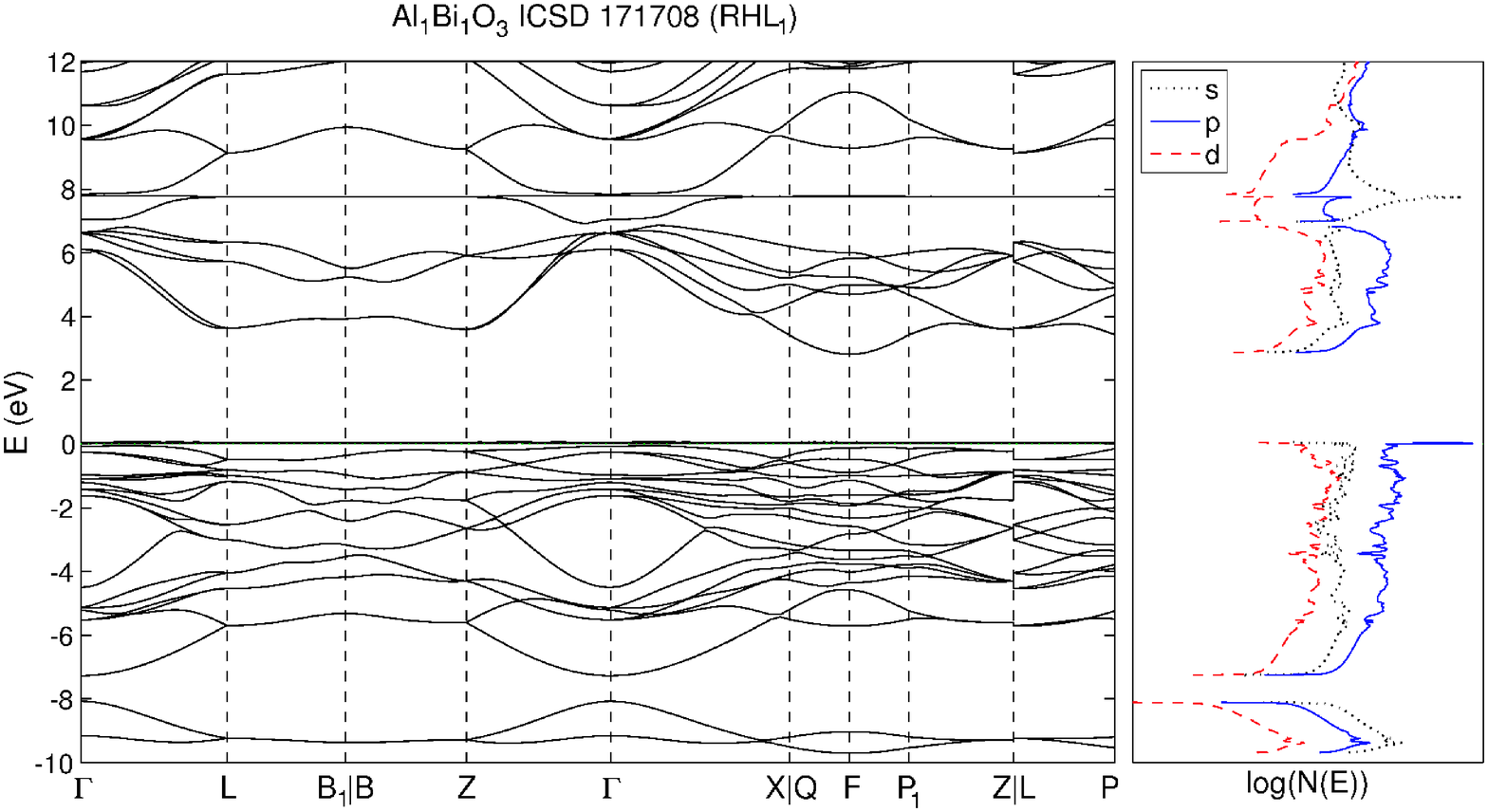
,height=42mm}}
\vspace{-3mm}
\caption{\small
Band structure of BiAlO$_3$ in RHL$_1$ lattice.}
\label{figbandRHL1}
\end{figure}
%
\begin{figure}[h!]
\centerline{\epsfig{file=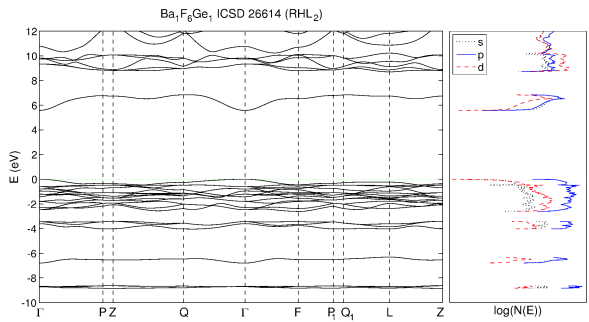
,height=42mm}}
\vspace{-3mm}
\caption{\small
Band structure of Ba(GeF$_6$) in RHL$_2$ lattice.}
\label{figbandRHL2}
\end{figure}

\begin{figure}[h!]
\centerline{\epsfig{file=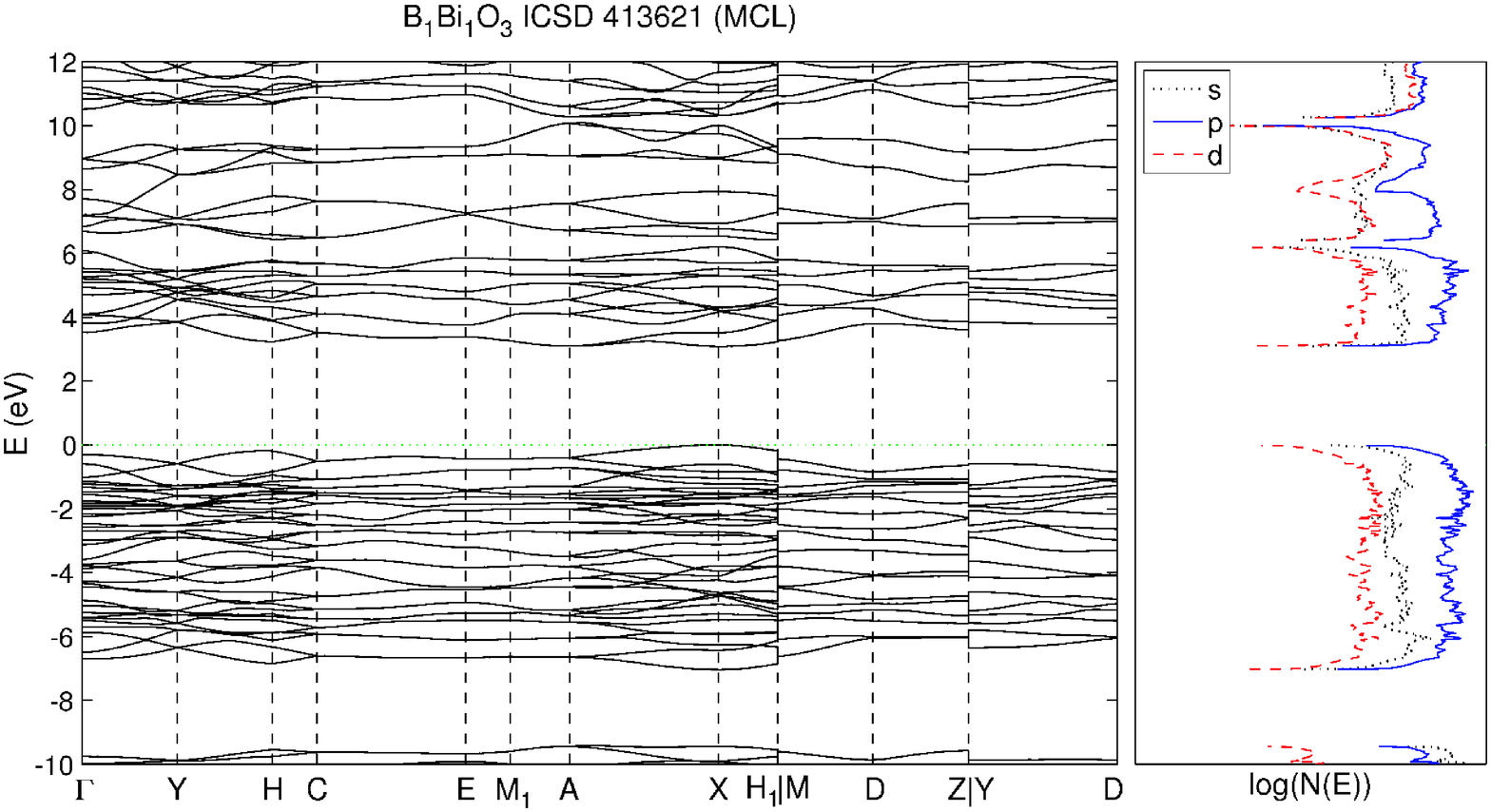
,height=42mm}}
\vspace{-3mm}
\caption{\small
Band structure of Bi(BO$_3$) in MCL lattice.}
\label{figbandMCL}
\end{figure}
%
\begin{figure}[h!]
\centerline{\epsfig{file=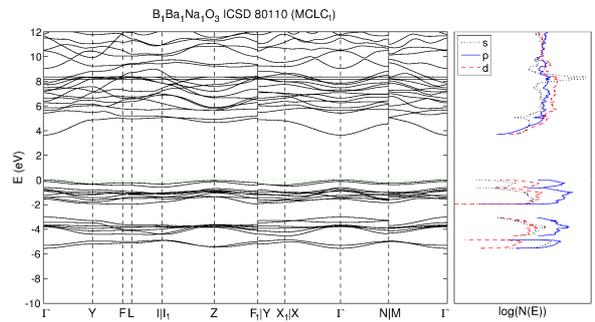
,height=42mm}}
\vspace{-3mm}
\caption{\small
Band structure of BaNa(BO$_3$) in MCLC$_1$ lattice.}
\label{figbandMCLC1}
\end{figure}
%
\begin{figure}[h!]
\centerline{\epsfig{file=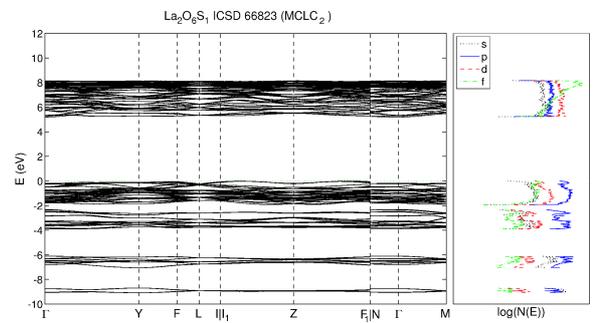
,height=42mm}}
\vspace{-3mm}
\caption{\small
Band structure of (LaO)$_2$(SO$_4$) in MCLC$_2$ lattice.}
\label{figbandMCLC2}
\end{figure}
%
\begin{figure}[h!]
\centerline{\epsfig{file=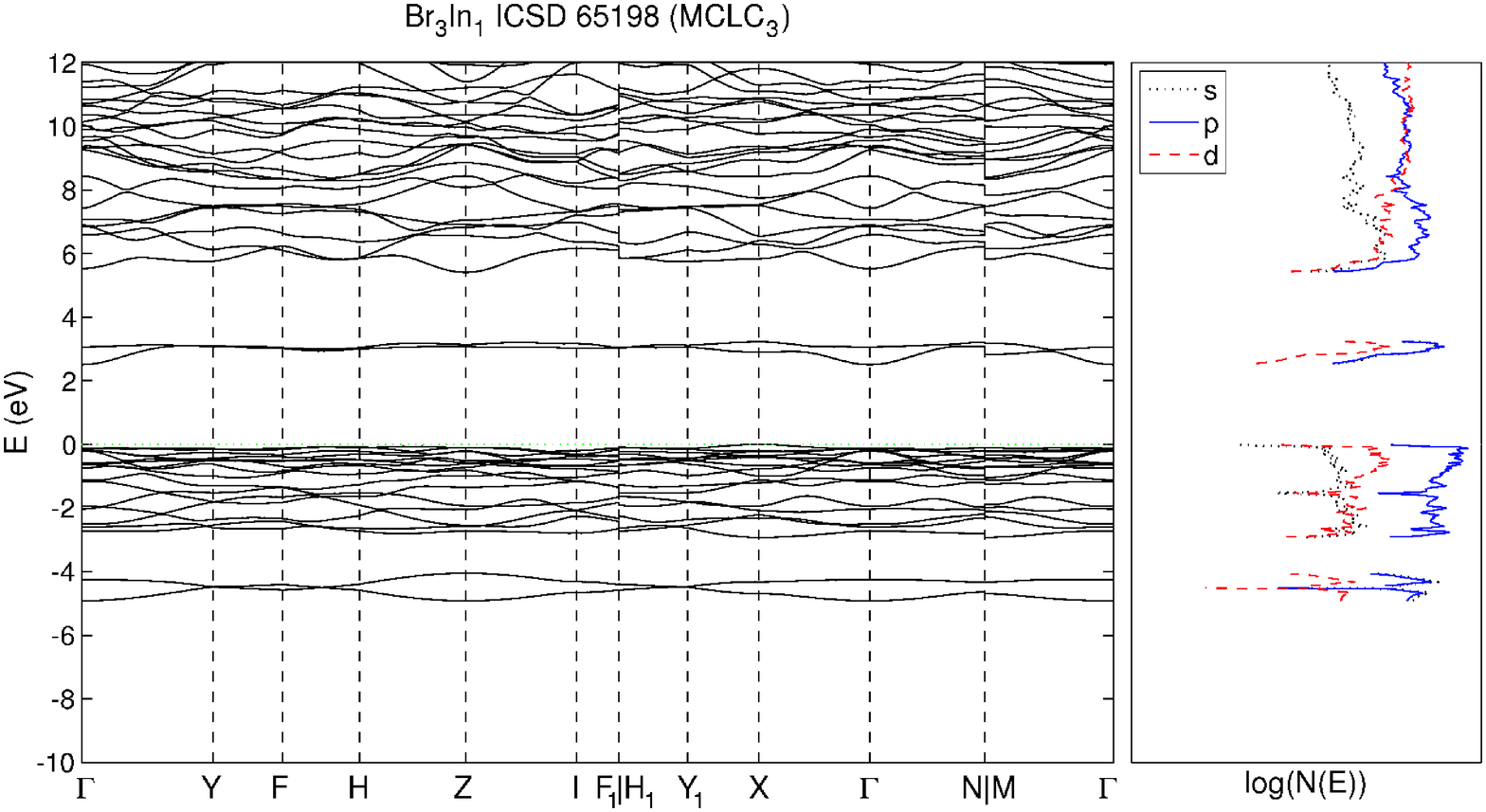
,height=42mm}}
\vspace{-3mm}
\caption{\small
Band structure of InBr$_3$ in MCLC$_3$ lattice.}
\label{figbandMCLC3}
\end{figure}
%
\clearpage 
\begin{figure}[h!]
\centerline{\epsfig{file=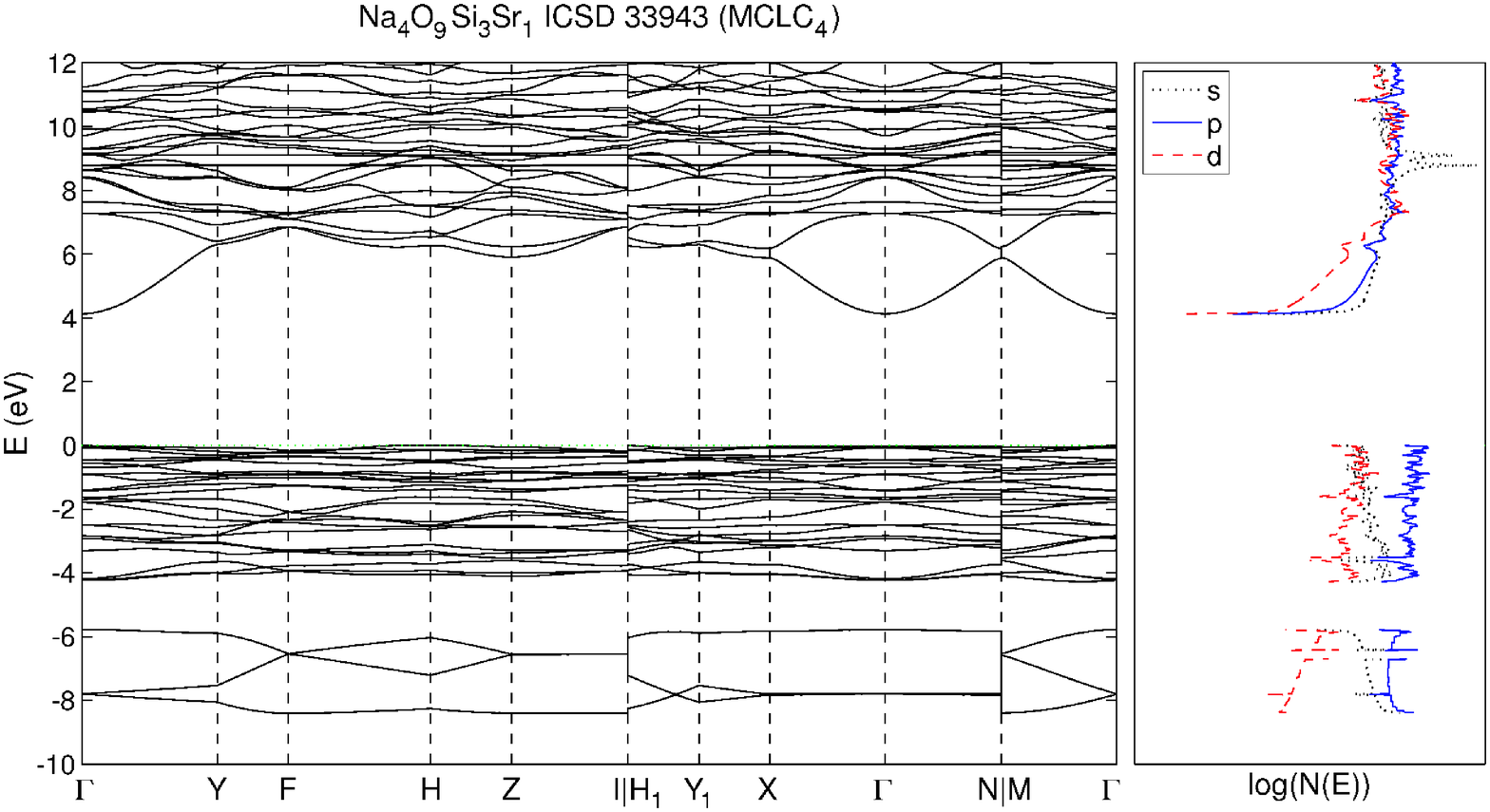
,height=42mm}}
\vspace{-3mm}
\caption{\small
Band structure of Na$_4$Sr(SiO$_3$)$_3$ in MCLC$_4$ lattice.}
\label{figbandMCLC4}
\end{figure}
%
\begin{figure}[h!]
\centerline{\epsfig{file=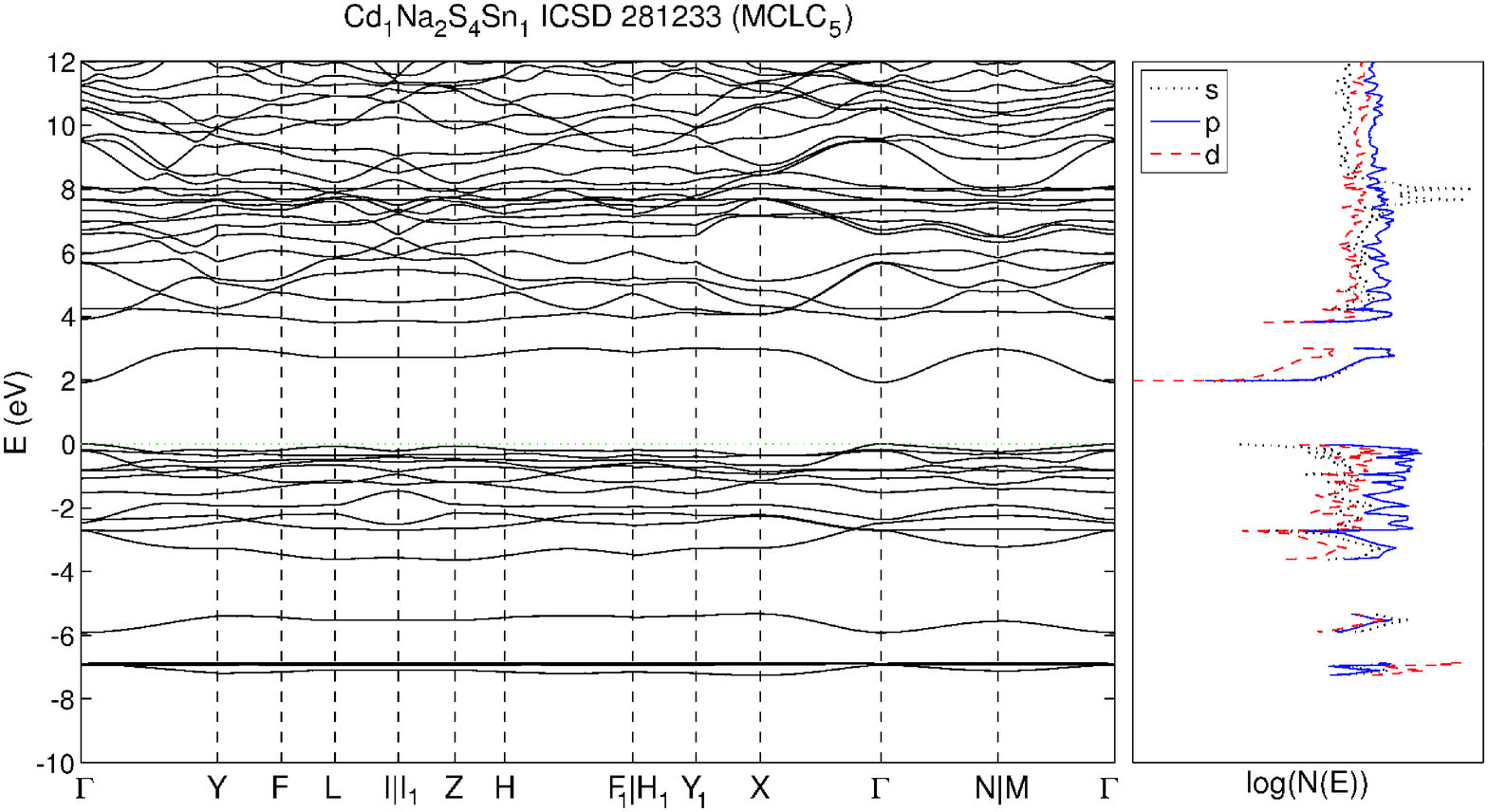
,height=42mm}}
\vspace{-3mm}
\caption{\small
Band structure of Na$_2$(CdSnS$_4$) in MCLC$_5$ lattice.}
\label{figbandMCLC5}
\end{figure}
%
\begin{figure}[h!]
\centerline{\epsfig{file=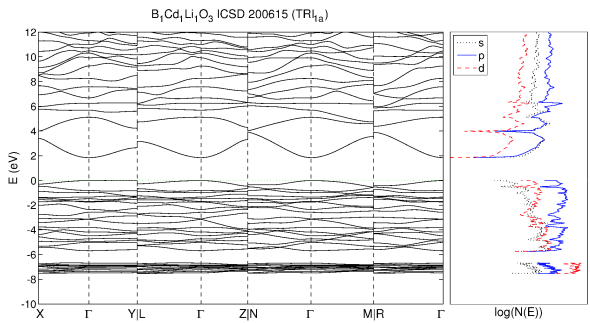
,height=42mm}}
\vspace{-3mm}
\caption{\small
Band structure of LiCd(BO$_3$) in TRI$_{1a}$ lattice.}
\label{figbandTRI1a}
\end{figure}
%
\begin{figure}[h!]
\centerline{\epsfig{file=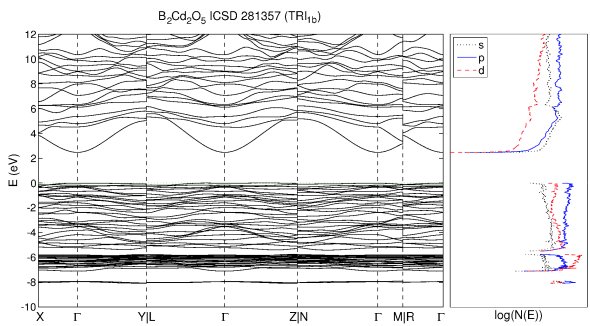,height=42mm}}
\vspace{-3mm}
\caption{\small
Band structure of Cd$_2$(B$_2$O$_5$) in TRI$_{1b}$ lattice.}
\label{figbandTRI1b}
\end{figure}

\section{Acknowledgments}
We thank Romain Gaume, Stephanie Lam, Robert Feigelson, Ohad Levy,
Mike Mehl, Gus Hart, Leeor Kronik, Roman Chepulskyy and Michal
Jahn\'atek for fruitful discussions.
 Research supported by ONR
(N00014-07-1-0878, N00014-07-1-1085, N00014-09-1-0921,
N00014-10-1-0436), and NSF (DMR-0639822, DMR-0650406). We are grateful
for extensive use of the Teragrid resources (MCA-07S005). SC
acknowledges the Feinberg support at the Weizmann Institute of
Science.


\begin{thebibliography}{57}
\expandafter\ifx\csname natexlab\endcsname\relax\def\natexlab#1{#1}\fi
\providecommand{\bibinfo}[2]{#2}
\ifx\xfnm\relax \def\xfnm[#1]{\unskip,\space#1}\fi
\bibitem[{Xiang et~al.(1995)Xiang, Sun, Briceno, Lou, Wang, Chang,
  Wallace-Freedman, Chen, and Schultz}]{Xiang06231995}
\bibinfo{author}{X.~D. Xiang}, \bibinfo{author}{X.~Sun},
  \bibinfo{author}{G.~Briceno}, \bibinfo{author}{Y.~Lou},
  \bibinfo{author}{K.-A. Wang}, \bibinfo{author}{H.~Chang},
  \bibinfo{author}{W.~G. Wallace-Freedman}, \bibinfo{author}{S.-W. Chen},
  \bibinfo{author}{P.~G. Schultz}, \bibinfo{journal}{Science}
  \bibinfo{volume}{268} (\bibinfo{year}{1995}) \bibinfo{pages}{1738--1740}.
\bibitem[{Koinuma and Takeuchi(2004)}]{koinuma_nmat_review2004}
\bibinfo{author}{H.~Koinuma}, \bibinfo{author}{I.~Takeuchi},
  \bibinfo{journal}{Nature Materials} \bibinfo{volume}{3}
  (\bibinfo{year}{2004}) \bibinfo{pages}{429--438}.
\bibitem[{Spivack et~al.(2003)Spivack, Cawse, Whisenhunt, Johnson, Shalyaev,
  Male, Pressman, Ofori, Soloveichik, Patel, Chuck, Smith, Jordan, Brennan,
  Kilmer, and Williams}]{Spivack20035}
\bibinfo{author}{J.~L. Spivack}, \bibinfo{author}{J.~N. Cawse},
  \bibinfo{author}{D.~W. Whisenhunt}, \bibinfo{author}{B.~F. Johnson},
  \bibinfo{author}{K.~V. Shalyaev}, \bibinfo{author}{J.~Male},
  \bibinfo{author}{E.~J. Pressman}, \bibinfo{author}{J.~Y. Ofori},
  \bibinfo{author}{G.~L. Soloveichik}, \bibinfo{author}{B.~P. Patel},
  \bibinfo{author}{T.~L. Chuck}, \bibinfo{author}{D.~J. Smith},
  \bibinfo{author}{T.~M. Jordan}, \bibinfo{author}{M.~R. Brennan},
  \bibinfo{author}{R.~J. Kilmer}, \bibinfo{author}{E.~D. Williams},
  \bibinfo{journal}{Applied Catalysis A: General} \bibinfo{volume}{254}
  (\bibinfo{year}{2003}) \bibinfo{pages}{5 -- 25}.
\bibitem[{Boussie et~al.(2003)Boussie, Diamond, Goh, Hall, LaPointe, Leclerc,
  Lund, Murphy, Shoemaker, Tracht, Turner, Zhang, Uno, Rosen, and
  Stevens}]{Boussie2003}
\bibinfo{author}{T.~R. Boussie}, \bibinfo{author}{G.~M. Diamond},
  \bibinfo{author}{C.~Goh}, \bibinfo{author}{K.~A. Hall},
  \bibinfo{author}{A.~M. LaPointe}, \bibinfo{author}{M.~Leclerc},
  \bibinfo{author}{C.~Lund}, \bibinfo{author}{V.~Murphy},
  \bibinfo{author}{J.~A.~W. Shoemaker}, \bibinfo{author}{U.~Tracht},
  \bibinfo{author}{H.~Turner}, \bibinfo{author}{J.~Zhang},
  \bibinfo{author}{T.~Uno}, \bibinfo{author}{R.~K. Rosen},
  \bibinfo{author}{J.~C. Stevens}, \bibinfo{journal}{J.\ Am.\ Chem.\ Soc.}
  \bibinfo{volume}{125} (\bibinfo{year}{2003}) \bibinfo{pages}{4306--4317}.
\bibitem[{Potyrailo et~al.(2003)Potyrailo, Chisholm, Morris, Cawse, Flanagan,
  Hassib, Molaison, Ezbiansky, Medford, and Reitz}]{Potyrailo2003}
\bibinfo{author}{R.~A. Potyrailo}, \bibinfo{author}{B.~J. Chisholm},
  \bibinfo{author}{W.~G. Morris}, \bibinfo{author}{J.~N. Cawse},
  \bibinfo{author}{W.~P. Flanagan}, \bibinfo{author}{L.~Hassib},
  \bibinfo{author}{C.~A. Molaison}, \bibinfo{author}{K.~Ezbiansky},
  \bibinfo{author}{G.~Medford}, \bibinfo{author}{H.~Reitz},
  \bibinfo{journal}{J. Comput. Chem.} \bibinfo{volume}{5}
  (\bibinfo{year}{2003}) \bibinfo{pages}{472--478}.
\bibitem[{Potyrailo and Takeuchi(2005)}]{Potyrailo2005}
\bibinfo{author}{R.~A. Potyrailo}, \bibinfo{author}{I.~Takeuchi},
  \bibinfo{journal}{Measurement Science and Technology} \bibinfo{volume}{16}
  (\bibinfo{year}{2005}) \bibinfo{pages}{1}.
\bibitem[{Chiang et~al.(1998)Chiang, Sadoway, Aydinol, Jang, Huang, and
  Ceder}]{ceder:nature_1998}
\bibinfo{author}{Y.-M. Chiang}, \bibinfo{author}{D.~R. Sadoway},
  \bibinfo{author}{M.~K. Aydinol}, \bibinfo{author}{Y.-I. Jang},
  \bibinfo{author}{B.~Huang}, \bibinfo{author}{G.~Ceder},
  \bibinfo{journal}{Nature} \bibinfo{volume}{392} (\bibinfo{year}{1998})
  \bibinfo{pages}{694}.
\bibitem[{J\'{o}hannesson et~al.(2002)J\'{o}hannesson, Bligaard, Ruban,
  Skriver, Jacobsen, and N{\o}rskov}]{Johann02}
\bibinfo{author}{G.~H. J\'{o}hannesson}, \bibinfo{author}{T.~Bligaard},
  \bibinfo{author}{A.~V. Ruban}, \bibinfo{author}{H.~L. Skriver},
  \bibinfo{author}{K.~W. Jacobsen}, \bibinfo{author}{J.~K. N{\o}rskov},
  \bibinfo{journal}{Phys.\ Rev.\ Lett.} \bibinfo{volume}{88}
  (\bibinfo{year}{2002}) \bibinfo{pages}{255506}.
\bibitem[{Stucke and Crespi(2003)}]{Stucke03}
\bibinfo{author}{D.~P. Stucke}, \bibinfo{author}{V.~H. Crespi},
  \bibinfo{journal}{Nano Lett.} \bibinfo{volume}{3} (\bibinfo{year}{2003})
  \bibinfo{pages}{1183}.
\bibitem[{Curtarolo et~al.(2003)Curtarolo, Morgan, Persson, Rodgers, and
  Ceder}]{curtarolo:prl_2003_datamining}
\bibinfo{author}{S.~Curtarolo}, \bibinfo{author}{D.~Morgan},
  \bibinfo{author}{K.~Persson}, \bibinfo{author}{J.~Rodgers},
  \bibinfo{author}{G.~Ceder}, \bibinfo{journal}{Phys.\ Rev.\ Lett.}
  \bibinfo{volume}{91} (\bibinfo{year}{2003}) \bibinfo{pages}{135503}.
\bibitem[{Morgan et~al.(2005)Morgan, Ceder, and
  Curtarolo}]{morgan:meas_2005_ht}
\bibinfo{author}{D.~Morgan}, \bibinfo{author}{G.~Ceder},
  \bibinfo{author}{S.~Curtarolo}, \bibinfo{journal}{Meas. Sci. Technolog.}
  \bibinfo{volume}{16} (\bibinfo{year}{2005}) \bibinfo{pages}{296}.
\bibitem[{Curtarolo et~al.(2005)Curtarolo, Morgan, and Ceder}]{monster}
\bibinfo{author}{S.~Curtarolo}, \bibinfo{author}{D.~Morgan},
  \bibinfo{author}{G.~Ceder}, \bibinfo{journal}{Calphad} \bibinfo{volume}{29}
  (\bibinfo{year}{2005}) \bibinfo{pages}{163}.
\bibitem[{Fischer et~al.(2006)Fischer, Tibbetts, Morgan, and Ceder}]{Fischer06}
\bibinfo{author}{C.~C. Fischer}, \bibinfo{author}{K.~J. Tibbetts},
  \bibinfo{author}{D.~Morgan}, \bibinfo{author}{G.~Ceder},
  \bibinfo{journal}{Nature Materials} \bibinfo{volume}{5}
  (\bibinfo{year}{2006}) \bibinfo{pages}{641}.
\bibitem[{Bligaard et~al.(2003)Bligaard, Johannesson, Ruban, Skriver, Jacobsen,
  and N{\o}rskov}]{johannesson:ref2}
\bibinfo{author}{T.~Bligaard}, \bibinfo{author}{G.~H. Johannesson},
  \bibinfo{author}{A.~V. Ruban}, \bibinfo{author}{H.~L. Skriver},
  \bibinfo{author}{K.~W. Jacobsen}, \bibinfo{author}{J.~K. N{\o}rskov},
  \bibinfo{journal}{Appl.\ Phys.\ Lett.} \bibinfo{volume}{83}
  (\bibinfo{year}{2003}) \bibinfo{pages}{4527}.
\bibitem[{Anderson et~al.(2006)Anderson, Blidgaard, A.~Kustov, Greeley,
  Johannessen, Christensen, and N{\o}rskov}]{johannesson:ref3}
\bibinfo{author}{M.~P. Anderson}, \bibinfo{author}{T.~Blidgaard},
  \bibinfo{author}{K.~E.~L. A.~Kustov}, \bibinfo{author}{J.~Greeley},
  \bibinfo{author}{T.~Johannessen}, \bibinfo{author}{C.~H. Christensen},
  \bibinfo{author}{J.~K. N{\o}rskov}, \bibinfo{journal}{J. Catal.}
  \bibinfo{volume}{239} (\bibinfo{year}{2006}) \bibinfo{pages}{501}.
\bibitem[{Levy et~al.(2010{\natexlab{a}})Levy, Chepulskii, Hart, and
  Curtarolo}]{curtarolo:art53}
\bibinfo{author}{O.~Levy}, \bibinfo{author}{R.~V. Chepulskii},
  \bibinfo{author}{G.~L.~W. Hart}, \bibinfo{author}{S.~Curtarolo},
  \bibinfo{journal}{J. Am. Chem. Soc.} \bibinfo{volume}{132}
  (\bibinfo{year}{2010}{\natexlab{a}}) \bibinfo{pages}{833}.
\bibitem[{Levy et~al.(2010{\natexlab{b}})Levy, Hart, and
  Curtarolo}]{curtarolo:art49}
\bibinfo{author}{O.~Levy}, \bibinfo{author}{G.~L.~W. Hart},
  \bibinfo{author}{S.~Curtarolo}, \bibinfo{journal}{J. Am. Chem. Soc.}
  \bibinfo{volume}{132} (\bibinfo{year}{2010}{\natexlab{a}}) \bibinfo{pages}{4830}.
\bibitem[{Kolmogorov and Curtarolo(2006)}]{curtarolo:art21}
\bibinfo{author}{A.~N. Kolmogorov}, \bibinfo{author}{S.~Curtarolo},
  \bibinfo{journal}{Phys.\ Rev.\ B} \bibinfo{volume}{73} (\bibinfo{year}{2006})
  \bibinfo{pages}{180501(R)}.
\bibitem[{Calandra et~al.(2007)Calandra, Kolmogorov, and
  Curtarolo}]{calandra:LiB_superconductivity_2007}
\bibinfo{author}{M.~Calandra}, \bibinfo{author}{A.~N. Kolmogorov},
  \bibinfo{author}{S.~Curtarolo}, \bibinfo{journal}{Phys.\ Rev.\ B}
  \bibinfo{volume}{75} (\bibinfo{year}{2007}) \bibinfo{pages}{144506}.
\bibitem[{Kolmogorov et~al.(2008)Kolmogorov, Calandra, and
  Curtarolo}]{kolmogorov:ternary_borides_LiB_2008}
\bibinfo{author}{A.~N. Kolmogorov}, \bibinfo{author}{M.~Calandra},
  \bibinfo{author}{S.~Curtarolo}, \bibinfo{journal}{Phys.\ Rev.\ B}
  \bibinfo{volume}{78} (\bibinfo{year}{2008}) \bibinfo{pages}{094520}.
\bibitem[{Wolverton et~al.(2008)Wolverton, Siegel, Akbarzadeh, and
  Ozolins}]{Wolverton2008}
\bibinfo{author}{C.~Wolverton}, \bibinfo{author}{D.~J. Siegel},
  \bibinfo{author}{A.~R. Akbarzadeh}, \bibinfo{author}{V.~Ozolins},
  \bibinfo{journal}{Journal of Physics: Condensed Matter} \bibinfo{volume}{20}
  (\bibinfo{year}{2008}) \bibinfo{pages}{064228}.
\bibitem[{Siegel et~al.(2007)Siegel, Wolverton, and
  Ozolins}]{Siegel_PhysRevB.76.134102}
\bibinfo{author}{D.~J. Siegel}, \bibinfo{author}{C.~Wolverton},
  \bibinfo{author}{V.~Ozolins}, \bibinfo{journal}{Phys. Rev. B}
  \bibinfo{volume}{76} (\bibinfo{year}{2007}) \bibinfo{pages}{134102}.
\bibitem[{Akbarzadeh et~al.(2007)Akbarzadeh, Ozolins, and
  Wolverton}]{Akbarzadeh2007}
\bibinfo{author}{A.~R. Akbarzadeh}, \bibinfo{author}{V.~Ozolins},
  \bibinfo{author}{C.~Wolverton}, \bibinfo{journal}{Advanced Materials}
  \bibinfo{volume}{19} (\bibinfo{year}{2007}) \bibinfo{pages}{3233--3239}.
\bibitem[{Massalski et~al.(1990)Massalski, Okamoto, Subramanian, and
  Kacprzak}]{Massalski}
\bibinfo{editor}{T.~B. Massalski}, \bibinfo{editor}{H.~Okamoto},
  \bibinfo{editor}{P.~R. Subramanian}, \bibinfo{editor}{L.~Kacprzak} (Eds.),
  \bibinfo{title}{Binary alloy phase diagrams}, \bibinfo{publisher}{American
  Society for Metals}, \bibinfo{address}{Materials Park, OH},
  \bibinfo{year}{1990}.
\bibitem[{Villars et~al.(2004)Villars, Berndt, Brandenburg, Cenzual, Daams,
  Hulliger, Massalski, Okamoto, Osaki, Prince, Putz, and Iwata}]{Pauling}
\bibinfo{author}{P.~Villars}, \bibinfo{author}{M.~Berndt},
  \bibinfo{author}{K.~Brandenburg}, \bibinfo{author}{K.~Cenzual},
  \bibinfo{author}{J.~Daams}, \bibinfo{author}{F.~Hulliger},
  \bibinfo{author}{T.~Massalski}, \bibinfo{author}{H.~Okamoto},
  \bibinfo{author}{K.~Osaki}, \bibinfo{author}{A.~Prince},
  \bibinfo{author}{H.~Putz}, \bibinfo{author}{S.~Iwata}, \bibinfo{journal}{J.
  Alloys Compound.} \bibinfo{volume}{367} (\bibinfo{year}{2004})
  \bibinfo{pages}{293--297}.
\bibitem[{Mighell and Karen(1993)}]{ICSD}
\bibinfo{author}{A.~D. Mighell}, \bibinfo{author}{V.~L. Karen},
  \bibinfo{journal}{Acta Cryst.} \bibinfo{volume}{A49} (\bibinfo{year}{1993})
  \bibinfo{pages}{c409}.
\bibitem[{Belsky et~al.(2002)Belsky, Hellenbrandt, Karen, and Luksch}]{ICSD3}
\bibinfo{author}{A.~Belsky}, \bibinfo{author}{M.~Hellenbrandt},
  \bibinfo{author}{V.~L. Karen}, \bibinfo{author}{P.~Luksch},
  \bibinfo{journal}{Acta Cryst.} \bibinfo{volume}{B58} (\bibinfo{year}{2002})
  \bibinfo{pages}{364--369}.
\bibitem[{NÃ¸rskov et~al.(2009)NÃ¸rskov, Bligaard, Rossmeisl, and
  Christensen}]{Norskov2009natchem}
\bibinfo{author}{J.~K. NÃ¸rskov}, \bibinfo{author}{T.~Bligaard},
  \bibinfo{author}{J.~Rossmeisl}, \bibinfo{author}{C.~H. Christensen},
  \bibinfo{journal}{Nature Chemistry} \bibinfo{volume}{1}
  (\bibinfo{year}{2009}) \bibinfo{pages}{37}.
\bibitem[{Hautier et~al.(2010)Hautier, Fischer, Jain, Mueller, and
  Ceder}]{Ceder_Chem_Materials}
\bibinfo{author}{G.~Hautier}, \bibinfo{author}{C.~Fischer},
  \bibinfo{author}{A.~Jain}, \bibinfo{author}{T.~Mueller},
  \bibinfo{author}{G.~Ceder}, \bibinfo{journal}{{\it Finding Nature\`s Missing
  Ternary Oxide Compounds using Machine Learning and Density Functional
  Theory}}  (\bibinfo{year}{2010}).
\bibitem[{Bradley and Cracknell(1972)}]{Bradley72}
\bibinfo{author}{C.~J. Bradley}, \bibinfo{author}{A.~P. Cracknell},
  \bibinfo{title}{The mathematical theory of symmetry in solids: Representation
  theory for point groups and space groups}, \bibinfo{publisher}{Clarendon
  Press}, \bibinfo{address}{Oxford}, \bibinfo{year}{1972}.
\bibitem[{Burns and Glazer(1990)}]{Burns90}
\bibinfo{author}{G.~Burns}, \bibinfo{author}{A.~M. Glazer},
  \bibinfo{title}{Space Groups for Solid State Scientists},
  \bibinfo{publisher}{Academic Press}, \bibinfo{address}{Boston},
  \bibinfo{year}{1990}.
\bibitem[{Miller and Love(1967)}]{Miller67}
\bibinfo{author}{S.~C. Miller}, \bibinfo{author}{W.~F. Love},
  \bibinfo{title}{Tables of Irreducible Representations of Space Groups and
  Co-Representations of Magnetic Groups}, \bibinfo{publisher}{Pruett Press},
  \bibinfo{address}{Boulder}, \bibinfo{year}{1967}.
\bibitem[{Kovalev(1965)}]{Kovalev65}
\bibinfo{author}{O.~V. Kovalev}, \bibinfo{title}{Irreducible Representations of
  the Space Groups}, \bibinfo{publisher}{Gordon and Breach},
  \bibinfo{address}{New York}, \bibinfo{year}{1965}.
\bibitem[{Casher et~al.(1969)Casher, Glucky, and Gur}]{Casher69}
\bibinfo{author}{A.~Casher}, \bibinfo{author}{M.~Glucky},
  \bibinfo{author}{Y.~Gur}, \bibinfo{title}{The Irreducible Representations of
  Space Groups}, \bibinfo{publisher}{W. A. Benjamin Inc.},
  \bibinfo{address}{New York}, \bibinfo{year}{1969}.
\bibitem[{Curtarolo et~al.(2009)Curtarolo, Hart, Setyawan, Mehl, Jahnatek,
  Chepulskii, Levy, and Morgan}]{AFLOW}
\bibinfo{author}{S.~Curtarolo}, \bibinfo{author}{G.~L.~W. Hart},
  \bibinfo{author}{W.~Setyawan}, \bibinfo{author}{M.~Mehl},
  \bibinfo{author}{M.~Jahnatek}, \bibinfo{author}{R.~V. Chepulskii},
  \bibinfo{author}{O.~Levy}, \bibinfo{author}{D.~Morgan},
  \bibinfo{journal}{{\it ``AFLOW: software for high-throughput calculation of
  material properties''}, {\sf http://materials.duke.edu/aflow.html}}
  (\bibinfo{year}{2009}).
\bibitem[{Giannozzi et~al.(2009)Giannozzi, Baroni, Bonini, Calandra, Car,
  Cavazzoni, Ceresoli, Chiarotti, Cococcioni, Dabo, {Dal Corso}, {de
  Gironcoli}, Fabris, Fratesi, Gebauer, Gerstmann, Gougoussis, Kokalj, Lazzeri,
  Martin-Samos, Marzari, Mauri, Mazzarello, Paolini, Pasquarello, Paulatto,
  Sbraccia, Scandolo, Sclauzero, Seitsonen, Smogunov, Umari, and
  Wentzcovitch}]{quantum_espresso_2009}
\bibinfo{author}{P.~Giannozzi}, \bibinfo{author}{S.~Baroni},
  \bibinfo{author}{N.~Bonini}, \bibinfo{author}{M.~Calandra},
  \bibinfo{author}{R.~Car}, \bibinfo{author}{C.~Cavazzoni},
  \bibinfo{author}{D.~Ceresoli}, \bibinfo{author}{G.~L. Chiarotti},
  \bibinfo{author}{M.~Cococcioni}, \bibinfo{author}{I.~Dabo},
  \bibinfo{author}{A.~{Dal Corso}}, \bibinfo{author}{S.~{de Gironcoli}},
  \bibinfo{author}{S.~Fabris}, \bibinfo{author}{G.~Fratesi},
  \bibinfo{author}{R.~Gebauer}, \bibinfo{author}{U.~Gerstmann},
  \bibinfo{author}{C.~Gougoussis}, \bibinfo{author}{A.~Kokalj},
  \bibinfo{author}{M.~Lazzeri}, \bibinfo{author}{L.~Martin-Samos},
  \bibinfo{author}{N.~Marzari}, \bibinfo{author}{F.~Mauri},
  \bibinfo{author}{R.~Mazzarello}, \bibinfo{author}{S.~Paolini},
  \bibinfo{author}{A.~Pasquarello}, \bibinfo{author}{L.~Paulatto},
  \bibinfo{author}{C.~Sbraccia}, \bibinfo{author}{S.~Scandolo},
  \bibinfo{author}{G.~Sclauzero}, \bibinfo{author}{A.~P. Seitsonen},
  \bibinfo{author}{A.~Smogunov}, \bibinfo{author}{P.~Umari},
  \bibinfo{author}{R.~M. Wentzcovitch}, \bibinfo{journal}{J.\ Phys.:\ Conden.\
  Matt.} \bibinfo{volume}{21} (\bibinfo{year}{2009}) \bibinfo{pages}{395502}.
\bibitem[{Nguyen and Stehl\'{e}(2009)}]{Nguyen2009_Minkowsky}
\bibinfo{author}{P.~Q. Nguyen}, \bibinfo{author}{D.~Stehl\'{e}},
  \bibinfo{journal}{ACM Trans. Algorithms} \bibinfo{volume}{5}
  (\bibinfo{year}{2009}) \bibinfo{pages}{1--48}.
\bibitem[{Cracknell (1973)}]{Cracknell73}
\bibinfo{author}{A. P. Cracknell},
  \bibinfo{journal}{J. Phys. C: Solid State Phys.} \bibinfo{volume}{6}
  (\bibinfo{year}{1973}) \bibinfo{pages}{826}.
\bibitem[{Setyawan et~al.(2009)Setyawan, Gaume, Feigelson, and
  Curtarolo}]{curtarolo:art46}
\bibinfo{author}{W.~Setyawan}, \bibinfo{author}{R.~M. Gaume},
  \bibinfo{author}{R.~S. Feigelson}, \bibinfo{author}{S.~Curtarolo},
  \bibinfo{journal}{IEEE Trans. Nucl. Sci.} \bibinfo{volume}{56}
  (\bibinfo{year}{2009}) \bibinfo{pages}{2989}.
\bibitem[{Karen and Hellenbrandt(2002)}]{ICSD1}
\bibinfo{author}{V.~L. Karen}, \bibinfo{author}{M.~Hellenbrandt},
  \bibinfo{journal}{Acta Cryst.} \bibinfo{volume}{A58} (\bibinfo{year}{2002})
  \bibinfo{pages}{c367}.
\bibitem[{Brown et~al.(2005)Brown, Abrahams, Berndt, Faber, Karen, Motherwell,
  Villars, Westbrook, and McMahon}]{ICSD2}
\bibinfo{author}{I.~D. Brown}, \bibinfo{author}{S.~C. Abrahams},
  \bibinfo{author}{M.~Berndt}, \bibinfo{author}{J.~Faber},
  \bibinfo{author}{V.~L. Karen}, \bibinfo{author}{W.~D.~S. Motherwell},
  \bibinfo{author}{P.~Villars}, \bibinfo{author}{J.~D. Westbrook},
  \bibinfo{author}{B.~McMahon}, \bibinfo{journal}{Acta Cryst.}
  \bibinfo{volume}{A61} (\bibinfo{year}{2005}) \bibinfo{pages}{575--580}.
\bibitem[{Kohn and Sham(1965)}]{DFT}
\bibinfo{author}{W.~Kohn}, \bibinfo{author}{L.~J. Sham},
  \bibinfo{journal}{Phys. Rev.} \bibinfo{volume}{140} (\bibinfo{year}{1965})
  \bibinfo{pages}{A1133}.
\bibitem[{Blochl(1994)}]{PAW}
\bibinfo{author}{P.~E. Blochl}, \bibinfo{journal}{Phys.\ Rev.\ B}
  \bibinfo{volume}{50} (\bibinfo{year}{1994}) \bibinfo{pages}{17953}.
\bibitem[{Perdew et~al.(1996)Perdew, Burke, and Ernzerhof}]{PBE}
\bibinfo{author}{J.~P. Perdew}, \bibinfo{author}{K.~Burke},
  \bibinfo{author}{M.~Ernzerhof}, \bibinfo{journal}{Phys.\ Rev.\ Lett.}
  \bibinfo{volume}{77} (\bibinfo{year}{1996}) \bibinfo{pages}{3865}.
\bibitem[{Monkhorst and Pack(1976)}]{MonkhorstPack}
\bibinfo{author}{H.~J. Monkhorst}, \bibinfo{author}{J.~D. Pack},
  \bibinfo{journal}{Phys.\ Rev.\ B} \bibinfo{volume}{13} (\bibinfo{year}{1976})
  \bibinfo{pages}{5188}.
\bibitem[{Setyawan et~al.(2009)Setyawan, Diehl, and
  Curtarolo}]{curtarolo:art44}
\bibinfo{author}{W.~Setyawan}, \bibinfo{author}{R.~D. Diehl},
  \bibinfo{author}{S.~Curtarolo}, \bibinfo{journal}{Phys.\ Rev.\ Lett.}
  \bibinfo{volume}{102} (\bibinfo{year}{2009}) \bibinfo{pages}{055501}.
\bibitem[{Diehl et~al.(2008)Diehl, Setyawan, and Curtarolo}]{curtarolo:art43}
\bibinfo{author}{R.~D. Diehl}, \bibinfo{author}{W.~Setyawan},
  \bibinfo{author}{S.~Curtarolo}, \bibinfo{journal}{J.\ Phys.:\ Conden.\ Matt.}
  \bibinfo{volume}{20} (\bibinfo{year}{2008}) \bibinfo{pages}{314007}.
\bibitem[{Aryasetiawan and Gunnarsson(1998)}]{GW}
\bibinfo{author}{F.~Aryasetiawan}, \bibinfo{author}{O.~Gunnarsson},
  \bibinfo{journal}{Rep. Prog. Phys.} \bibinfo{volume}{61}
  (\bibinfo{year}{1998}) \bibinfo{pages}{237}.
\bibitem[{Duradev et~al.(1998)Duradev, Botton, Savrasov, Humphreys, and
  Sutton}]{DuradevDFTU}
\bibinfo{author}{S.~L. Duradev}, \bibinfo{author}{G.~A. Botton},
  \bibinfo{author}{S.~Y. Savrasov}, \bibinfo{author}{C.~J. Humphreys},
  \bibinfo{author}{A.~P. Sutton}, \bibinfo{journal}{Phys.\ Rev.\ B}
  \bibinfo{volume}{57} (\bibinfo{year}{1998}) \bibinfo{pages}{1505}.
\bibitem[{Liechtenstein et~al.(1995)Liechtenstein, Anisimov, and
  Zaanen}]{LiechDFTU}
\bibinfo{author}{A.~I. Liechtenstein}, \bibinfo{author}{V.~I. Anisimov},
  \bibinfo{author}{J.~Zaanen}, \bibinfo{journal}{Phys.\ Rev.\ B}
  \bibinfo{volume}{52} (\bibinfo{year}{1995}) \bibinfo{pages}{R5467}.
\bibitem[{{Leiria Campo Jr} and Cococcioni(2010)}]{Cococcioni_JPCM}
\bibinfo{author}{V.~{Leiria Campo Jr}}, \bibinfo{author}{M.~Cococcioni},
  \bibinfo{journal}{Journal of Physics: Condensed Matter} \bibinfo{volume}{22}
  (\bibinfo{year}{2010}) \bibinfo{pages}{055602}.
\bibitem[{Lang et~al.(1981)Lang, Baer, and Cox}]{Lang81}
\bibinfo{author}{J.~K. Lang}, \bibinfo{author}{Y.~Baer}, \bibinfo{author}{P.~A.
  Cox}, \bibinfo{journal}{J. Phys. F: Met. Phys.} \bibinfo{volume}{11}
  (\bibinfo{year}{1981}) \bibinfo{pages}{121}.
\bibitem[{Wegner et~al.(2006)Wegner, Bauer, Koroteev, Bihlmayer, Chulkov,
  Echenique, and Kaindl}]{LaUJ}
\bibinfo{author}{D.~Wegner}, \bibinfo{author}{A.~Bauer}, \bibinfo{author}{Y.~M.
  Koroteev}, \bibinfo{author}{G.~Bihlmayer}, \bibinfo{author}{E.~V. Chulkov},
  \bibinfo{author}{P.~M. Echenique}, \bibinfo{author}{G.~Kaindl},
  \bibinfo{journal}{Phys.\ Rev.\ B} \bibinfo{volume}{73} (\bibinfo{year}{2006})
  \bibinfo{pages}{115403}.
\bibitem[{Jiang et~al.(2005)Jiang, Adams, and van Schilfgaarde}]{CeUJ}
\bibinfo{author}{Y.~Jiang}, \bibinfo{author}{J.~B. Adams},
  \bibinfo{author}{M.~van Schilfgaarde}, \bibinfo{journal}{J.\ Chem.\ Phys.}
  \bibinfo{volume}{123} (\bibinfo{year}{2005}) \bibinfo{pages}{064701}.
\bibitem[{Harmon et~al.(1995)Harmon, Antropov, Liechtenstein, Solovyev, and
  Anisimov}]{GdUJ}
\bibinfo{author}{B.~N. Harmon}, \bibinfo{author}{V.~P. Antropov},
  \bibinfo{author}{A.~I. Liechtenstein}, \bibinfo{author}{I.~V. Solovyev},
  \bibinfo{author}{V.~I. Anisimov}, \bibinfo{journal}{J. Phys. Chem. Solids}
  \bibinfo{volume}{56} (\bibinfo{year}{1995}) \bibinfo{pages}{1521}.
\bibitem[{Harima(2010)}]{PrUJ}
\bibinfo{author}{H.~Harima}, \bibinfo{journal}{J. Mag. Mag. Mat.}
  \bibinfo{volume}{226} (\bibinfo{year}{2010}) \bibinfo{pages}{83}.
\bibitem[{Antonov et~al.(2001)Antonov, Harmon, and Yaresko}]{TmUJ}
\bibinfo{author}{V.~N. Antonov}, \bibinfo{author}{B.~N. Harmon},
  \bibinfo{author}{A.~N. Yaresko}, \bibinfo{journal}{Phys.\ Rev.\ B}
  \bibinfo{volume}{63} (\bibinfo{year}{2001}) \bibinfo{pages}{205112}.
\bibitem[{Jeong(2006)}]{YbUJ}
\bibinfo{author}{T.~Jeong}, \bibinfo{journal}{J.\ Phys.:\ Conden.\ Matt.}
  \bibinfo{volume}{18} (\bibinfo{year}{2006}) \bibinfo{pages}{6769}.

\end{thebibliography}
\end{document}